\documentclass[a4paper,10.5pt]{article}

\usepackage[dvips]{graphicx}
\usepackage{float}
\usepackage{epsfig,subfigure}
\usepackage{longtable}
\usepackage{rotating}
\usepackage{amsmath}
\usepackage{amsfonts}
\usepackage{amssymb}
\usepackage{times}
\usepackage{cite}
\usepackage{dcolumn}
\usepackage{bm}
\usepackage{hyperref}   
\usepackage{multirow}

\usepackage{color}

\newcommand{\sqrts}{\sqrt{s}}
\newcommand{\sqrtsnn}{\sqrt{s_{_{\mbox{\rm \tiny{NN}}}}}}
\newcommand{\RAA}{R_{_{\rm AA}}}

\newcommand{\pt}{p_{\rm T}}

\newcommand\cN{{\cal N}}
\newcommand\bR{{\cal B}}

\newcommand{\LumiInt}{\mathcal{L}_{\mbox{\rm \tiny{int}}}}

\newcommand{\pp}{{\rm{p-p}}}

\newcommand{\pPb}{{\rm{p-Pb}}}
\newcommand{\PbPb}{{\rm{Pb-Pb}}}
\newcommand{\mcfm}{{\sc mcfm}}

\newcommand{\bbbar}    {\mathit{b\,\bar{b}}}
\newcommand{\ttbar}    {t\bar{t}}

\newcommand{\MET}{\ensuremath{{E\!\!\!/}_{_{\rm T}}}}

\def\ttt#1{\texttt{\scriptsize #1}}

\begin{document}

\begin{center}
\vbox to 1 truecm {}

{\Large \bf Top-quark production in proton-nucleus and nucleus-nucleus}
\\[0.2cm] 
{\Large \bf collisions at LHC energies and beyond} \\[0.8cm]

{\large David~d'Enterria$^{1}$, Kriszti\'an Krajcz\'ar$^1$, and Hannu Paukkunen$^{2,3}$}\\[0.5cm]

{\it $^1$ CERN, PH Department, 1211 Geneva, Switzerland}\\[0.1cm]
{\it $^2$ Department of Physics, University of Jyv\"askyl\"a, P.O. Box 35, FI-40014 University of Jyv\"askyl\"a, Finland}\\[0.1cm]
{\it $^3$ Helsinki Institute of Physics, P.O. Box 64, FI-00014 University of Helsinki, Finland}

\vskip 3 truemm

\end{center}


\begin{abstract}
\noindent
Single and pair top-quark production in proton-lead (\pPb) and lead-lead (\PbPb) collisions at the 
CERN Large Hadron Collider (LHC) and future circular collider (FCC) energies, are studied with next-to-leading-order 
perturbative QCD calculations including nuclear parton distribution functions. 
At the LHC, the pair-production cross sections amount to $\sigma_{\ttbar}$~=~3.4~$\mu$b 
in \PbPb\ at $\sqrtsnn$~=~5.5~TeV, and $\sigma_{\ttbar}$~=~60~nb in \pPb\ at $\sqrtsnn$~=~8.8~TeV. 
At the FCC energies of $\sqrtsnn$~=~39 and 63~TeV, the same cross sections are factors of 90 and 55 times larger respectively. 
In the leptonic final-state $\ttbar\to W^+b\,W^-\bar{b}\to \bbbar\,\ell\ell\,\nu\nu$, after typical
acceptance and efficiency cuts, one expects about 90 and 300 top-quarks per nominal LHC-year and
4.7$\cdot10^4$ and $10^5$ per FCC-year in \PbPb\ and \pPb\ collisions respectively. 
The total $\ttbar$ cross sections, 
dominated by gluon fusion processes, are enhanced by 3--8\% in nuclear compared to \pp\ collisions due to an
overall net gluon antishadowing, although different regions of their differential distributions are depleted due to 
shadowing or EMC-effect corrections. The rapidity distributions of the decay leptons in $\ttbar$ processes can be used to
reduce the uncertainty on the Pb gluon density at high virtualities by up to 30\% at the LHC (full heavy-ion programme),
and by 70\% per FCC-year. The cross sections for single-top production 
in electroweak processes are also computed, yielding 
about a factor of 30 smaller number of measurable top-quarks after cuts, per system and per year. 
\end{abstract}



\section{Introduction}
\label{sec:0}

The multi-TeV energies available 
at the CERN Large Hadron Collider (LHC) have opened up the possibility to measure, for the first time in heavy-ion collisions,
various large-mass elementary particles. After the first observations
of the $W$~\cite{Chatrchyan:2012nt,Aad:2014bha} and $Z$~\cite{Chatrchyan:2011ua,Aad:2012ew} bosons, as well as
bottom-quark ($b$-jets)~\cite{Chatrchyan:2013exa}, there remains only three Standard Model (SM) elementary particles to be
directly measured in nucleus-nucleus collisions: the $\tau$ lepton, the Higgs boson, and the top quark. 
Whereas the $\tau$ measurement should be straightforward, 
that of the Higgs boson is beyond the LHC reach as it requires much larger
cross sections and/or luminosities~\cite{DdE}, 
such as those reachable at the proposed future circular collider (FCC)~\cite{Armesto:2014iaa} with about 
seven times larger center-of-mass energies than at the LHC. The study presented here shows, for the first time, 
that the top-quark --the heaviest elementary particle known-- will be produced (singly or in pairs) in
sufficiently large numbers to be observed in lead-lead (\PbPb) and proton-lead (\pPb) collisions at the LHC
and FCC.

Since the width of the top-quark ($\Gamma_{\rm t}\approx$~2~GeV) is much larger than the
parton-to-hadron transition scale given by $\Lambda_{_{\rm QCD}}\approx$~0.2~GeV, the top-quark is the only
coloured particle that decays before its hadronization. Its short lifetime, 
$\tau_0 = \hbar/\Gamma_{\rm t}\approx$~0.1~fm/c, implies that the top decays --into a
$t\to W\,b$ final-state with a nearly 100\% branching ratio~\cite{PDG}-- 
mostly\footnote{The typical transverse momentum of the produced top quark is usually smaller than 
its mass, $\pt < m_t$, and the Lorentz-boost factor is $\gamma \approx \cosh(y_t)$,
where $y_t$ is the $t$-quark rapidity. At the LHC ($|y_t|<3$) the Lorentz-dilated mean
decay time is $\tau = \gamma \tau_0 \approx 0.1-1 \, {\rm fm/c}$, and at
the FCC ($|y_t|<5$), $\tau = \gamma \tau_0 \approx 0.1-7.5 \, {\rm fm/c}$;
to be compared with the typical QGP formation time of $1$~fm/c, and 
lifetime of $10-20$~fm/c.}
within the strongly-interacting medium, such as the quark-gluon plasma
(QGP), formed in nuclear collisions. 
The large top-quark mass ($m_{\rm t}\approx$~173~GeV) provides a hard scale for high-accuracy
perturbative calculations of its quantum-chromodynamics (QCD) and electroweak production cross sections (next-to-next-to-leading-order, or NNLO, is the current theoretical state-of-the-art~\cite{Czakon:2012pz,Kidonakis:2012rm,Czakon:2013goa}).
At hadron colliders, top quarks are produced either in pairs, dominantly through the strong interaction, or
singly through the weak interaction. At the energies considered here, the dominant production channels, as
obtained at NLO accuracy with the \mcfm\ code~\cite{mcfm}, are: (i) gluon-gluon fusion, $g\,g\to\ttbar+X$,
contributing by 80--95\% to the total pair production (the remaining 5--20\% issuing from quark-antiquark
annihilation), (ii) $t$-channel single-top electroweak production $q\,b\to q'\,t+X$ (the $s$-channel process,
decreasing with energy, amounts to 5--1.5\% of the total single-$t$ cross section), and 
(iii) associated top plus $W$-boson, $g\,b\to W\,t+X$, production (increasing with energy, it amounts to
25--50\% of the $t$-channel process).

The theoretical motivations for a dedicated experimental measurement of the top-quark in heavy-ion collisions
are varied and include, at least, the following studies:
\begin{description}
\item (i) Constraints on nuclear parton distribution functions (nPDFs). The $\ttbar$ cross sections in
proton-proton (\pp) collisions can be used to constrain the proton PDFs~\cite{Czakon:2013tha}. In the
heavy-ion case, top-pair production probes the nuclear gluon density in an unexplored kinematic regime around
Bjorken-$x$ values, $x\approx 2\,m_{\rm t}/\sqrtsnn\approx$~5$\cdot$10$^{-3}$--0.05, and virtualities  
$Q^2\approx m_{\rm t}^2 \approx 3\cdot 10^4$~GeV$^2$, a region characterized by net positive, albeit small,
anti-shadowing corrections. In addition, at the FCC, the $b$-quark nPDF (in single-top production), and even the
top-quark nPDF itself, 
are generated dynamically by the constituent gluons and become necessary ingredients of the theoretical cross section calculations.
\item (ii) Heavy-quark energy loss dynamics. The top quark can radiate gluons before its $W\,b$ decay which,
  given its very-short lifetime, occurs mostly inside the QGP. Medium-induced gluon radiation off light-quarks and
  gluons, leading to ``jet quenching''~\cite{d'Enterria:2009am}, results in a factor of two reduction
  of jet yields in \PbPb\ compared to \pp\ collisions at $\sqrtsnn$~=~2.76~TeV~\cite{Chatrchyan:2011sx,Aad:2010bu}. 
  Although solid theoretical expectations for heavy-quark radiation predict a reduced amount of
  gluonstrahlung at small angles due to the ``dead cone'' effect~\cite{deadcone}, the experimental data 
  somehow unexpectedly shows the same amount of suppression for jets from light-flavours and $b$-quarks~\cite{Chatrchyan:2013exa}.
  The relative role of elastic and radiative scatterings on the energy loss of heavy-quarks 
  is an open issue in the field~\cite{Gossiaux:2012th,Djordjevic:2013xoa}. 
  The detailed study of top-quark production in heavy-ion collisions would therefore provide novel
  interesting insights on the mechanisms of parton energy loss.
  In addition, the study of boosted top-pairs
  (with transverse momenta above $\pt\approx$~1~TeV) traversing the QGP as a colour-singlet object for a fraction of their time,
  will allow one to probe the medium opacity at different space-time scales.
\item (iii) Colour reconnection in the QGP. The top mass, featuring the strongest coupling to the
Higgs field, is a fundamental SM parameter with far-reaching implications including the
stability of the electroweak vacuum~\cite{Alekhin:2012py}.
Currently, the dominant $m_{\rm t}$ systematic uncertainty  is of theoretical nature and connected to the modeling
of the colour connection and QCD interferences between the $\ttbar$ production and decay stages, and among the 
hadronic decay products. Indeed, the colour-flow (through gluon exchanges and/or
non-perturbative string overlaps) between the $t$ and $\bar{t}$ quarks, their decayed $b$-quarks, and 
the underlying event from multi-parton interactions and beam-remnants surrounding the initial hard 
scattering~\cite{top_colour_reco}, 
results in uncertainties on the reconstructed $m_{\rm t}$ of a few hundred MeV. The
amount of top quark interactions with the colour fields stretched among many partons involved in nuclear 
collisions will be obviously enhanced compared to more elementary 
systems. Thus, the reconstruction
of the top-quark mass in the QGP (assuming its feasibility is not jeopardized by the large $b$-quark energy loss
already observed in the data), or in proton-nucleus interactions, would provide interesting insights in
non-perturbative QCD effects on a crucial SM parameter. 
\end{description}

In this paper we mostly focus on nPDF constraints through top-pair production in \pPb\ 
and \PbPb\ collisions. 
We also provide the expected 
$\pt$ reach of top quark spectra at various energies to indicate where boosted final-states can be measured
for energy loss studies. The paper is organized as follows. In Section~\ref{sec:2} the theoretical setup used
is outlined, which is then used to compute the NLO cross sections at the LHC and FCC, and associated yields
expected after typical acceptance and efficiency cuts, for top-pair and single-top production, 
presented in Section~\ref{sec:3}. Section~\ref{sec:4} quantifies the impact on the nuclear PDFs
provided by the measurement of the rapidity distributions
of the decay leptons from top-quark pairs produced at the LHC and FCC, 
using a Hessian PDF reweighting technique~\cite{Paukkunen:2013grz,Paukkunen:2014zia}.
The main conclusions of the work are summarized in Section~\ref{sec:5}. 

\section{Theoretical setup} 
\label{sec:2}

The top-pair and single-top cross sections are computed at NLO accuracy with the \mcfm\
code~\cite{mcfm} (version 6.7) using the NLO CT10 proton PDFs (including its 52 eigenvector sets)~\cite{Lai:2010vv}
corrected for nuclear effects (shadowing, antishadowing and EMC)~\cite{Arneodo:1992wf}
through the EPS09 nPDFs (including its 30 error sets) for the Pb ion~\cite{Eskola:2009uj}.
As our main purpose is to provide estimates for the feasibility of different top-quark
measurements in nuclear collisions, we do not discuss here the (subleading) sensitivity of the cross sections
to different sets of proton PDFs (nor associated variations of the strong coupling $\alpha_s$). Also, while there are 
other nuclear PDF sets available~\cite{Hirai:2007sx,Schienbein:2009kk,deFlorian:2011fp},
we only employ EPS09 here as it is the only nPDF set that is consistent with the dijet
measurements in \pPb\ collisions at the LHC~\cite{Chatrchyan:2014hqa,Paukkunen:2014pha}
(the data would probably agree also with the latest nPDFs by nCTEQ~\cite{Kusina:2014wwa}, but
these sets are not available at the time of writing this article).
We run the following \mcfm\ processes: $\ttt{141}$ for $\ttbar$ production, 
$\ttt{161,166}$ for single-(anti)top in the $t$-channel, $\ttt{171,176}$ in the s-channel, and $\ttt{181,186}$
for associated $t\,W$ production. 
We note that, at NLO, the theoretical processes defining $t\,W$ production partially overlap with those
contributing to top-quark pair production~\cite{Campbell:2005bb}. This is accounted for in our \mcfm\
$t\,W$ cross sections calculations by vetoing the additional emission of a $b$-jet.
The code also
properly accounts for the different isospin ($u$- and $d$-quark) content of the
Pb nucleus, 
which has a small impact on the electroweak single-top processes.

All numerical results have been obtained using the latest SM parameters for particle masses,
widths and couplings~\cite{PDG}, and fixing the default renormalization and factorization scales at 
$\mu = \mu_F = \mu_R$~=~$m_{\rm t}$ for $\ttbar$ and $t$-,$s$-channel single-top, and at $\mu = \mu_F =
\mu_R$~=~$p_{_{\rm T,min;b-jet}} = 50$~GeV for the $t\,W$ processes. 
The NLO calculations used here reproduce well the cross sections experimentally measured 
at the LHC in \pp\ collisions at $\sqrts$~=~7,~8~TeV 
for $\ttbar$~\cite{Khachatryan:2010ez,Chatrchyan:2011nb,Chatrchyan:2011ew,Chatrchyan:2012vs,Chatrchyan:2012bra,Chatrchyan:2012saa,Chatrchyan:2013kff,Chatrchyan:2013faa,Aad:2010ey,Aad:2011yb,Aad:2012qf,ATLAS:2012aa,Khachatryan:2014iya,Aad:2012mza,Aad:2012hg}, 
$t$-channel single-top~\cite{Chatrchyan:2011vp,Chatrchyan:2012ep,Khachatryan:2014iya,Aad:2012ux,Aad:2014fwa}, 
and associated $t\,W$~\cite{Chatrchyan:2012zca,Chatrchyan:2014tua,Aad:2012xca} production.
Incorporation of next-to-NLO corrections~\cite{Czakon:2012pz} would increase the theoretical cross sections,
by about 10\%, i.e. the so-called $K$-factor amounts to $K = \sigma_{_{\rm NNLO}}/\sigma_{_{\rm NLO}}\approx$~1.10, 
and further improve the data-theory agreement. 
The computed nucleon-nucleon cross sections are scaled by the Pb mass number ($A$~=~208) to obtain the \pPb\
cross sections, and by $A^2$~=~43\,264 in the \PbPb\ case, as expected for hard scattering processes in nuclear
collisions. 
The uncertainties of the theoretical cross sections are obtained 
from the values computed using the eigenvector sets of, first, the CT10 and, then, EPS09 PDFs and adding them in quadrature,
as well as by independently varying the renormalization and factorization $\mu_F$ and $\mu_R$ scales within a
factor of two (for the central CT10 and EPS09 set). The PDF and scale uncertainties amount each to
3--10\% for $\ttbar$ and 
3--6\% for single top.
The scale uncertainties have no impact on the nPDF constraints derived below given that 
the scale dependence of the nuclear modifications in PDFs around $\mu=m_{\rm t}$ is very mild (see
e.g. Fig.~1 in Ref.~\cite{Eskola:2012rg}) and that differences between proton and nuclear PDFs are obtained via
ratios of (\pPb,\PbPb)/(\pp) cross sections at the same colliding energy, where those mostly cancel out.


\section{Top-pair and single-top cross sections and yields}
\label{sec:3}

\begin{figure}[htpb]
\centering
\epsfig{figure=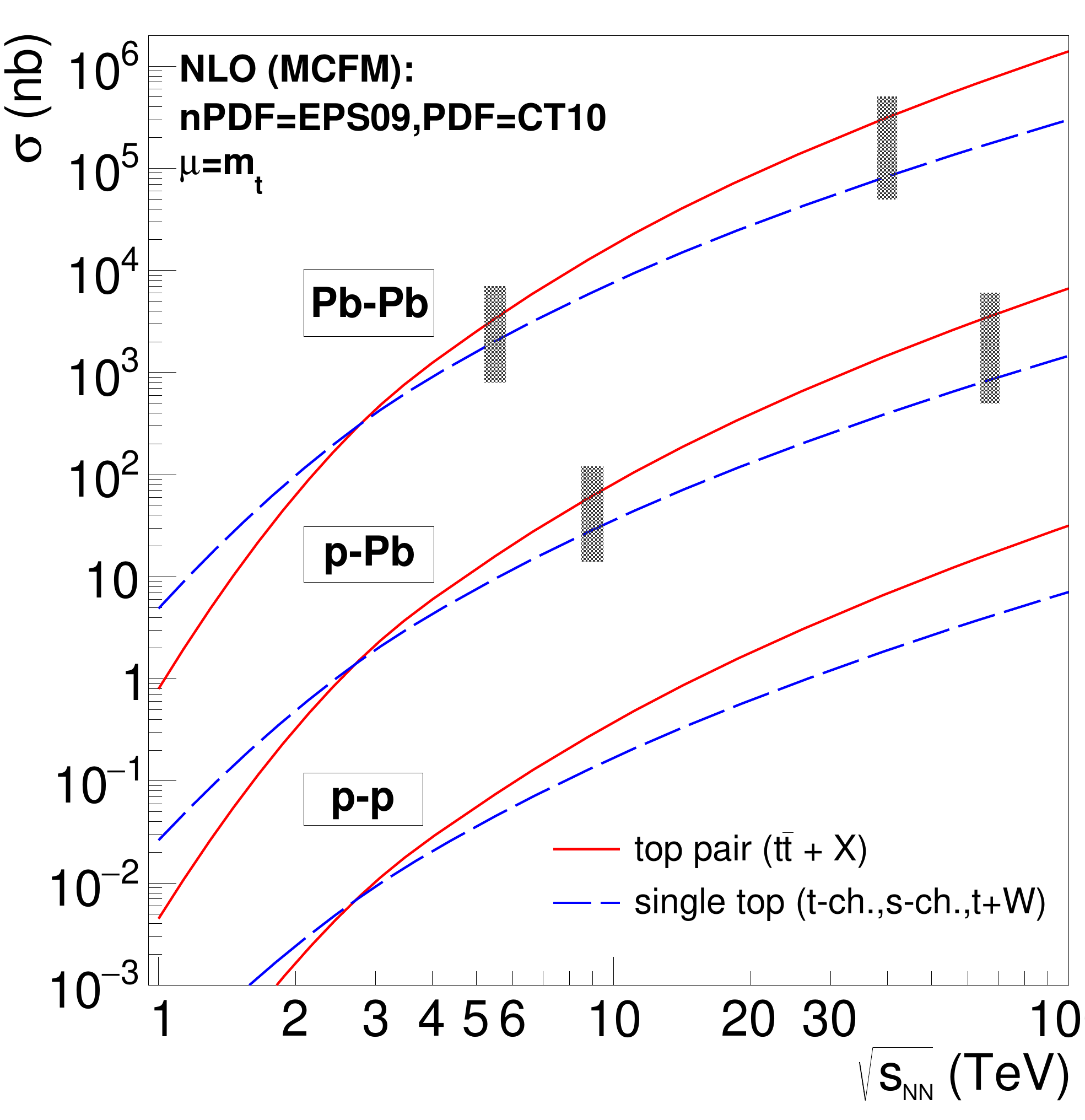,width=0.60\columnwidth}
\caption{Total cross sections for top pair and single-top (sum of $t$-,$s$-channels plus
  $t\,W$ processes) production in \PbPb\, \pPb\ and \pp\ collisions as a function of c.m. energy. The dashed
  boxes indicate the nominal nucleon-nucleon c.m. energies, $\sqrtsnn$~=~5.5, 8.8, 39, 63~TeV, of the
  heavy-ion runs at the LHC and FCC.\label{fig:1}}
\end{figure}

Table~\ref{tab:1} collects the total cross sections, and associated scale and PDF uncertainties, for top-pair
and single-top production in \pPb\ and \PbPb\ collisions at LHC and FCC energies, obtained as described in the
previous Section. In the case of $\ttbar$ and $t\,W$ production, a net EPS09 gluon antishadowing in collisions with Pb
ions results in an increase of the total production cross sections by about 2--8\% compared to those
($A$-scaled) obtained for \pp\ collisions using the proton CT10 PDF. For $t$- and $s$-channel single-top cross sections,
the overall nuclear modifications are quite insignificant, $\pm$2\% depending on the energy. 
Figure~\ref{fig:1} shows the total top-pair and single-top cross sections as a function of collision energy 
for \pp, \pPb\ and \PbPb\ collisions in the range of c.m. energies in the nucleon-nucleon system of $\sqrtsnn
\approx$~~1--100~TeV. The single-top curves are obtained adding the $t$-,$s$-channel and $t\,W$ cross sections
listed in Table~\ref{tab:1}. In general, top-quark pair production is a factor of two (four) larger than the sum
of single top processes at the LHC (FCC).
The nominal LHC and FCC energies for \pPb\ and \PbPb\ collisions are indicated by dashed
boxes in the plot. Going from LHC to FCC, the total cross sections increase by significant factors,
$\times$(55--90) for $\ttbar$ and $\times$(30--40) for single-top.

\begin{table}[htbp]
\caption{Inclusive cross sections for top-pair and single-top ($t$-channel, $s$-channel, and $t\,W$)
production in \PbPb\ and \pPb\ collisions at LHC and FCC energies, obtained at NLO accuracy with \mcfm.
The first uncertainty is due to theoretical scale variations, and the second one to the CT10 and EPS09 PDF
errors added in quadrature.
\label{tab:1}}
\vspace{-0.25cm}
\begin{center}
\tabcolsep=1.1mm
\begin{tabular}{l|c|c|c|c}\hline\hline
\hspace{0mm} System / Process: \hspace{0mm} & \hspace{0mm} top pair ($\ttbar$) \hspace{0mm} & \hspace{0mm} single-top ($t$-channel) \hspace{0mm} & 
\hspace{0mm} single-top ($s$-channel) \hspace{-1.0mm} & \hspace{-1.0mm} single-top ($t\,W$) \hspace{0mm} \\
\hspace{0mm} (\mcfm\ process) \hspace{0mm} & \hspace{0mm}  (\ttt{141}) \hspace{0mm} & 
\hspace{0mm} (\ttt{161,166})\hspace{0mm} & \hspace{0mm} (\ttt{171,176}) \hspace{0mm} & 
\hspace{0mm} (\ttt{181,186}) \hspace{0mm}\\\hline

\hspace{0mm} \PbPb\ $\sqrtsnn$~=~5.5~TeV \hspace{0mm} & 
\hspace{0mm} 3.40$\pm$0.42$\pm$0.37 $\mu$b \hspace{0mm} & \hspace{0mm} 1.61$\pm$0.05$\pm$0.08 $\mu$b \hspace{0mm} & 
\hspace{0mm} 110$\pm$4$\pm$6 nb \hspace{0mm} & \hspace{0mm} 313$\pm$13$\pm$41 nb \hspace{0mm} \\
\hspace{0mm} \pPb\;\, $\sqrtsnn$~=~8.8~TeV \hspace{0mm} & 
\hspace{0mm} 58.8$\pm$7.1$\pm$3.8 nb \hspace{0mm} & \hspace{0mm} 21.1$\pm$0.63$\pm$0.63 nb \hspace{0mm} & 
\hspace{0mm} 1.09$\pm$0.03$\pm$0.04 nb \hspace{0mm} & \hspace{0mm} 5.26$\pm$0.21$\pm$0.37 nb \hspace{0mm} \\
\hline

\hspace{0mm} \PbPb\ $\sqrtsnn$~=~39~TeV \hspace{0mm} & 
\hspace{0mm} 302$\pm$33$\pm$12 $\mu$b \hspace{0mm} & \hspace{0mm} 54.6$\pm$1.6$\pm$2.2 $\mu$b \hspace{0mm} & 
\hspace{0mm} 1.31$\pm$0.05$\pm$0.08 $\mu$b \hspace{0mm} & \hspace{0mm} 24.2$\pm$1.6$\pm$1.3 $\mu$b \hspace{0mm} \\
\hspace{0mm} \pPb\;\, $\sqrtsnn$~=~63~TeV \hspace{0mm} & 
\hspace{0mm} 3.20$\pm$0.35$\pm$0.10 $\mu$b \hspace{0mm} & \hspace{0mm} 518$\pm$16$\pm$17 nb \hspace{0mm} & 
\hspace{0mm} 10.9$\pm$0.5$\pm$0.5 nb \hspace{0mm} & \hspace{0mm} 246$\pm$24$\pm$11 nb \hspace{0mm} \\

\hline\hline

\end{tabular}
\end{center}
\end{table}

\begin{figure}[htpb]
\centering
\epsfig{figure=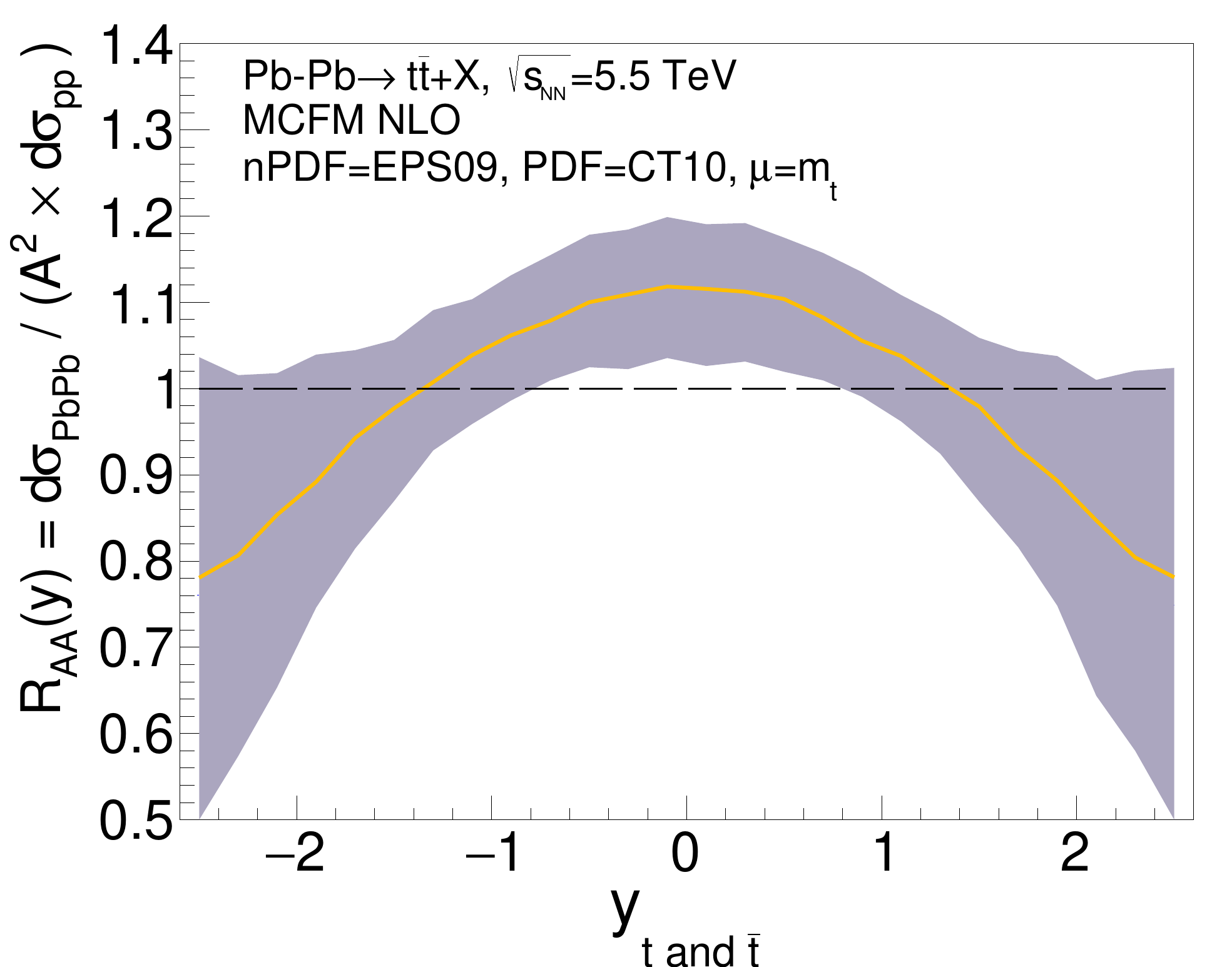,width=0.45\columnwidth}
\epsfig{figure=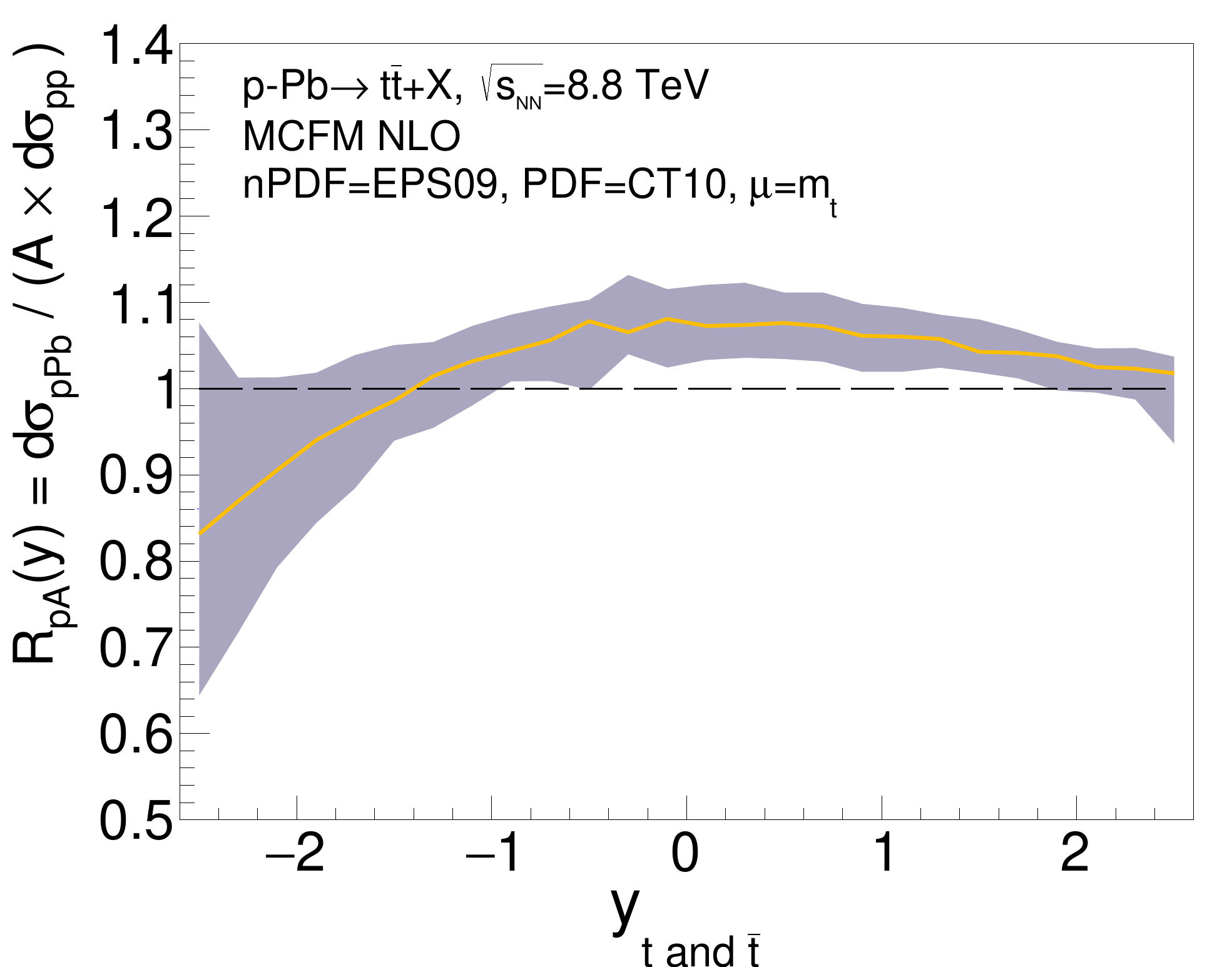,width=0.45\columnwidth}
\epsfig{figure=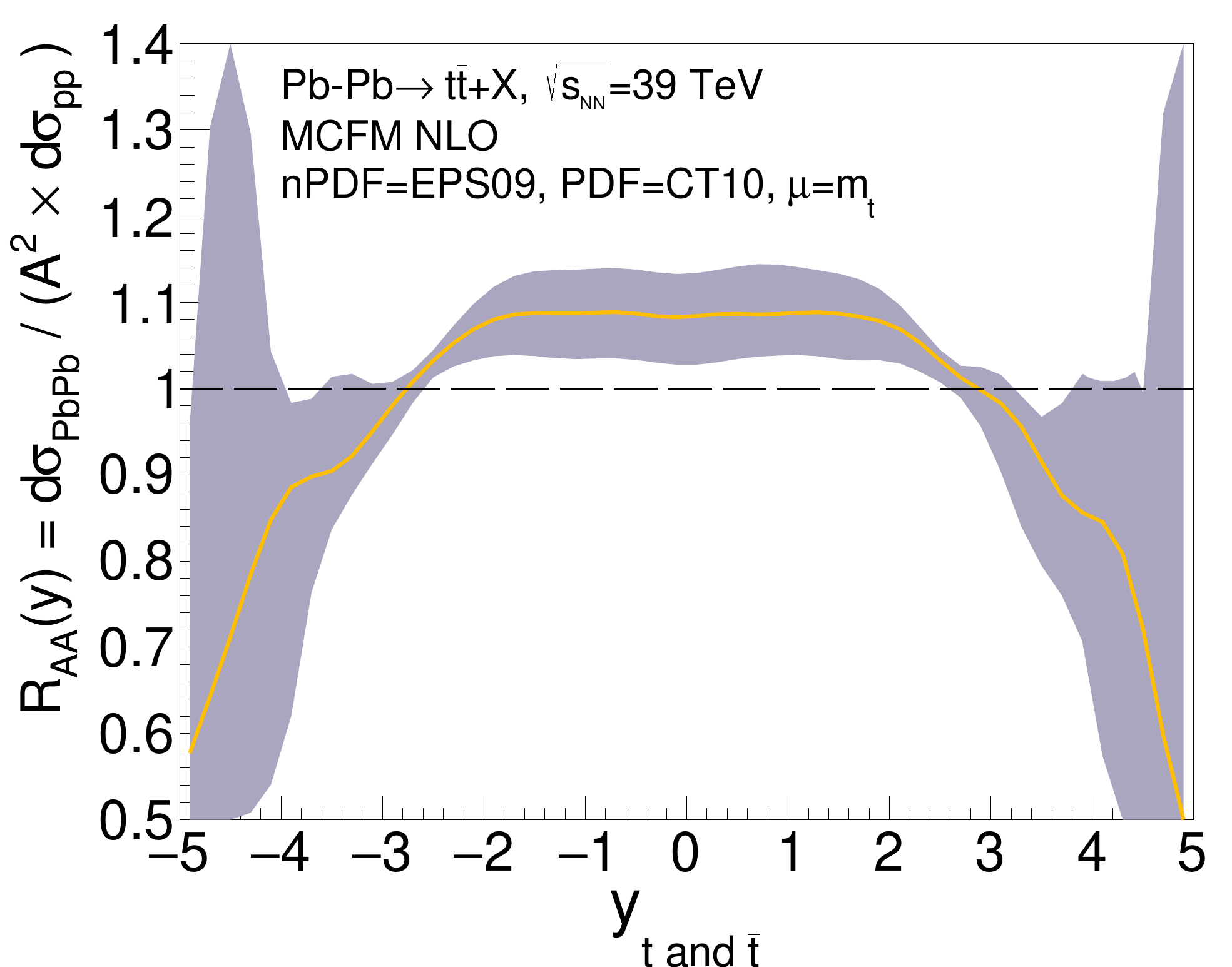,width=0.45\columnwidth}
\epsfig{figure=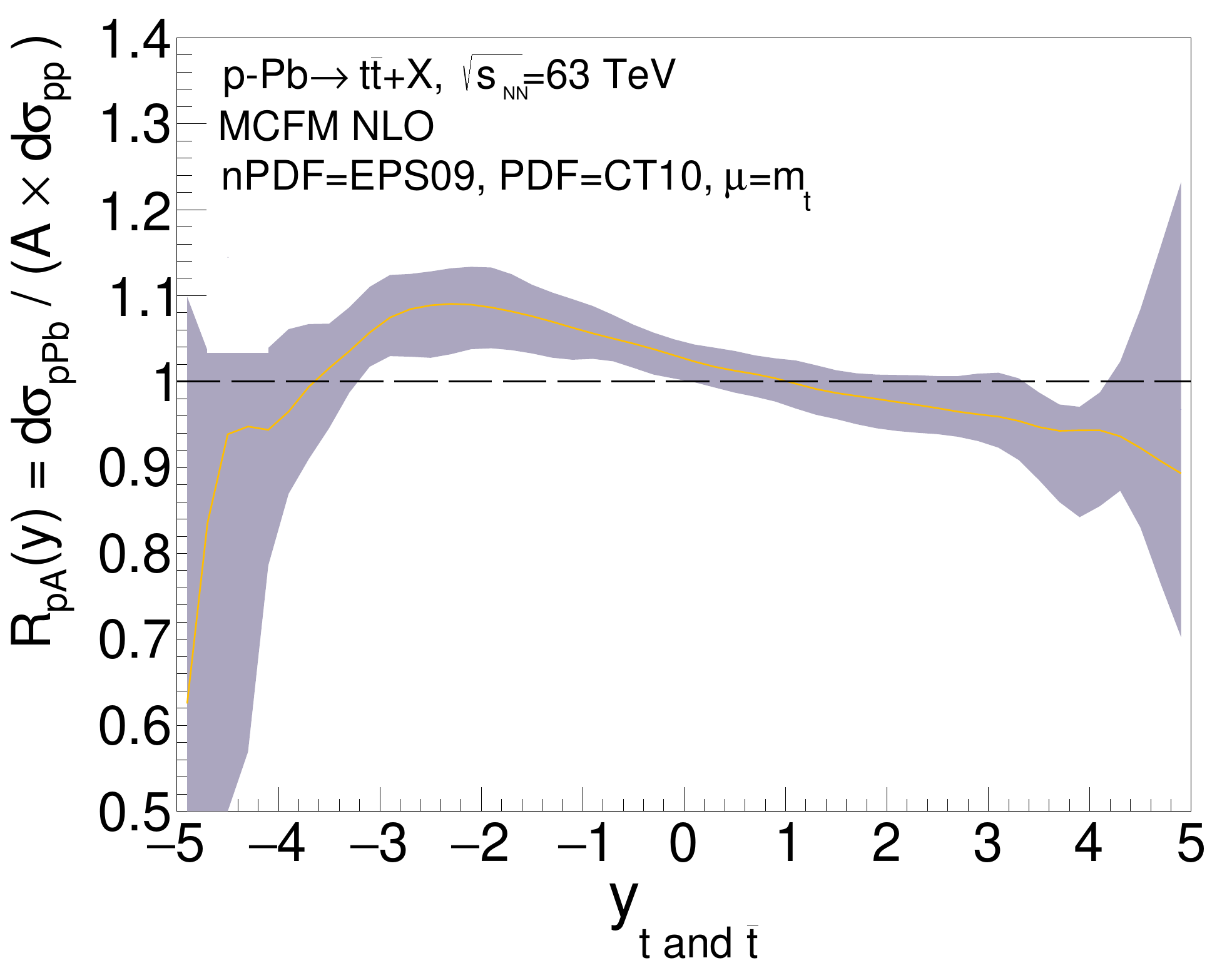,width=0.45\columnwidth}
\caption{Theoretical nuclear modification factor 
as a function of rapidity of the individual top and antitop quarks in $\ttbar$ production 
in \PbPb\ (left panels) and \pPb\ (right panels) at $\sqrtsnn$~=~5.5, 8.8~TeV (LHC, top panels) and
39, 63~TeV (FCC, bottom panels), computed at NLO accuracy with \mcfm. The central curve indicates the result
obtained with central EPS09 parametrization and the grey band the corresponding nPDF uncertainty. 
\label{fig:dsigmady_ttbar}}
\end{figure}

The impact of nuclear PDF modifications on the yields for a given hard process is usually quantified through
the nuclear modification factor $\RAA$ given by the ratio of cross sections in nuclear over proton-proton
collisions scaled by $A$ or $A^2$. The theoretical $\RAA(y_{\rm t,\bar{t}})$ factors as a function of the rapidity
of the produced top and antitop quarks are shown in Fig.~\ref{fig:dsigmady_ttbar} for $\ttbar$ 
at LHC (top panels) and FCC (bottom panels) energies. The central
curves indicate the result obtained with the central EPS09 set and the grey bands show the corresponding nPDF
uncertainties. All the results presented in those, and the following, plots are given in the center-of-mass frame of the colliding species.
In general, the $\RAA(y_{\rm t,\bar{t}})$ distributions reveal very similar trends for $\ttbar$
and single-top (not shown here) processes, although the $t$-channel and $s$-channel processes have a factor of
two smaller uncertainties, as expected, given that single-top production is dominated by quark-induced
processes whose densities in the nucleus are better known than the gluon ones which produce most of the
top pairs. At the LHC, for both (single and pair) production mechanisms, the nPDF effects increase the average top-quark distributions at
central rapidities by about 10\% (antishadowing) while they deplete them by 20\% at backward rapidities (also
in the forward direction in \PbPb) due to the so-called EMC effect at large-$x$~\cite{Arneodo:1992wf} (see
also Fig.~\ref{fig:data1} later). At the FCC, the higher collision energies as well as the larger kinematical
coverage assumed for the detectors at this future facility give access to smaller momentum fractions $x$ where
EPS09 predicts moderate shadowing even at high virtualities $Q\sim m_t$. This leads to additional suppression at
forward rapidities (and also in the backward direction in \PbPb).


The cross sections listed in Table~\ref{tab:1} and plotted in Fig.~\ref{fig:1} 
are total inclusive ones and do not include the $t\to
W\,b$ decays, nor any experimental acceptance/analysis requirements on the final-state particles. 
The determination of the expected yields at the LHC and FCC requires accounting for top and $W$-boson 
decays plus acceptance and reconstruction efficiency losses.
The $W$ leptonic branching fractions, $W\to\ell^{\pm} \nu_{\ell}$ (with $\ell$ = e, $\mu$, $\tau$), amount to 1/9 for each
lepton flavour, the other 2/3 being due to $W$ dijet (quark-antiquark) decays. In this work we will only consider leptonic
$W$ decays characterized by a final-state with an isolated electron or muon plus missing transverse energy ($\MET$) 
from the neutrino, because the $W$ dijet-decays are much more difficult to reconstruct in the large background
of heavy-ion collisions, and also potentially subject to parton energy loss effects (although they could be 
certainly tried in the ``cleaner'' \pPb\ environment). For the top-pairs measurement, $\ttbar\to
W^+b\,W^-\bar{b}\to \bbbar\,\ell\ell\,\nu\nu$, the combination of electron and muon decays for both $W$-bosons
($ee,\mu\mu,e\mu,\mu e$) reduces the total cross sections by a factor\footnote{Including also $e^\pm$ and
$\mu^\pm$ from leptonic tau-decays in the $\ttbar\to e\tau,\mu\tau,\tau\tau + \MET$ final-states would
decrease the corresponding branching ratio only to $\bR_{_{\rm \ell\ell}} \approx$~1/16.} of 
$\bR_{_{\rm \ell\ell}} = 4/9^2\approx$~1/20. 
For the single-top case, we will only consider the yields for associated $t\,W$ production which shares a
characteristic final-state signature, $t\,W \to W\,b\,W \to b\,\ell\ell\,\nu\nu$, very similar to that of
top-pair production. Indeed, the experimental observation of $t$-channel (let alone the much more suppressed
$s$-channel) single-top, with one less charged lepton and neutrino, is much more challenging on top of the
expected large $W,Z$+jets background (in \PbPb, at least, although it should be feasible in \pPb\ collisions).

In order to compute the expected number of top-quarks measurable at the LHC and FCC, 
we include in the \mcfm\ generator-level calculations the typical analysis and fiducial requirements for
$b$-jets, charged leptons, and missing transverse energy from the unidentified neutrinos, used in
similar \pp\ measurements at the LHC~\cite{Chatrchyan:2011nb,Chatrchyan:2012bra,Chatrchyan:2014tua}. Although 
some of these \pp\ experimental requirements may seem optimistic for the more complex environment encountered
in heavy-ion collisions, they are validated by future experimental projections of the CMS collaboration~\cite{CMS:2013gga}.
In the case of FCC, we extend the pseudorapidity coverage for charged-lepton tracking and 
$b$-jet secondary vertexing from the LHC range of $|\eta|$~=~2.5, to $|\eta|$~=~5. The details of all
selection criteria are given in Table~\ref{tab:cuts}. 
We reconstruct the $b$-jets with the anti-$k_{_{\rm T}}$ jet clustering algorithm~\cite{Cacciari:2008gp} with
distance parameter R~=~0.5, and we require the high-$\pt$ charged lepton to be separated from the closest
$b$-jet within an ($\eta,\phi$) isolation radius of R$_{_{\rm isol}}$ = 0.3. 

\begin{table}[htbp]
\caption{List of analysis cuts on single $\pt$ and $\eta$ for $b$-jets and isolated leptons 
$\ell = e^{\pm},\mu^{\pm}$, and on the neutrinos $\MET$, typically employed in
top-pair ($\ttbar\to \bbbar\,\ell\ell\,\nu\nu$)~\cite{Chatrchyan:2011nb,Chatrchyan:2012bra} 
and single-top plus $W$-boson ($t\,W \to b\,\ell\ell\,\nu\nu$)~\cite{Chatrchyan:2014tua}
measurements in fully-leptonic final-states in \pp\ collisions at the LHC, applied in our generator-level
studies. 
\label{tab:cuts}}
\vspace{-0.25cm}
\begin{center}
\begin{tabular}{l}\hline\hline
\hspace{0mm}  Analysis cuts  \\\hline
$b$-jets (anti-k$_{_{\rm T}}$ algorithm with R~=~0.5): $\pt > 30$ GeV; $|\eta| <$ 2.5 (LHC), 5 (FCC)\\ 
charged leptons $\ell$ (R$_{_{\rm isol}}$ = 0.3): $\pt > 20$ GeV; $|\eta| <$ 2.5 (LHC), 5 (FCC)\\
neutrinos: $\MET > 40$ GeV \\\hline\hline 
\end{tabular}
\end{center}
\end{table}

The combination of all analysis cuts listed in Table~\ref{tab:cuts} results in total acceptances of order
${\cal A}_{_{\ttbar}}\approx{\cal A}_{_{tW}}\approx$~40\% (50\%) for $\ttbar$ and $t\,W$ measurements at the LHC (FCC).
In addition, one has to account for experimental $b$-jet tagging efficiencies, which we conservatively take of
the order of 50\% as determined in \PbPb\ collisions at $\sqrts$~=~2.76~TeV~\cite{Chatrchyan:2013exa}. For single-top,
this results in an extra $\varepsilon_{_{tW}}\approx$~0.5 reduction of the measured yields. In the $\ttbar$
case, in order to tag the event as such, one usually only requires a single $b$-jet (out of the two produced) to
be identified, and thus the associated efficiency is larger: $\varepsilon_{_{\ttbar}} \approx 1 - (1 - 0.5)^2 = 0.75$. 
The combination of acceptance, analysis requirements, and efficiency losses results in an overall efficiency factor of 
${\cal A}_{_{\ttbar}}\times\varepsilon_{_{\ttbar}} \approx$~30\% (40\%) for the 
final $\ttbar$ yields\footnote{We note that, in the \PbPb\ case, parton energy loss effects which can
bring the $b$-jet below the $\pt$ threshold criterion ($\pt>$~30~GeV) and result in an additional inefficiency to tag the
$\ttbar$ event, are unlikely to affect {\it both} $b$-jets at the same time. Indeed, for simple geometrical
reasons if one top-quark is produced and decays through the denser region, the other one emitted back-to-back will
go through a thinner medium layer and its associated $b$-jet will be tagged with our considered probability.}
at the LHC (FCC). The same factor for $t\,W$ production is 
${\cal A}_{_{tW}}\times\varepsilon_{_{tW}} \approx$~20\% (25\%) at the LHC (FCC).
Possible backgrounds, mostly from $W,Z$+jets, $WZ$, and $ZZ$ production sharing similar final-state signatures as
both top-quark production processes, can be minimized by applying dedicated jet-veto requirements and/or
extra criteria on the invariant masses of the two high-$\pt$ leptons, e.g. away from the $Z$ boson peak 
($|m_{_{Z}}- m_{\ell\ell}|>$~15~GeV). We do not directly compute the impact of
such backgrounds here as the applied analysis cuts 
already reduce them 
to a manageable level according to the existing \pp\ measurements. In particular, the application of
$m_{\ell\ell}$ cuts would reduce the visible yields by an extra 10\% which is, however, compensated by the
fact that our NLO calculations need to be scaled by about the same amount to match the current \pp\ data (and NNLO predictions).
We note also that, despite the larger hadronic backgrounds in nuclear collisions, the instantaneous
luminosities in the heavy-ion operation mode at the LHC (and FCC) result in a very small event pileup,
at variance with the \pp\ case, and make the top-quark measurements accessible without the complications from
overlapping nuclear collisions occurring simultaneously in the same bunch crossing.

The expected total number of top-quarks (adding the $t$ and $\bar{t}$ values) produced in \PbPb\ and \pPb\ in one
year at the nominal luminosities for each colliding system, obtained via 
$\cN = \sigma \cdot\bR_{_{\rm \ell\ell}} \cdot\LumiInt\cdot{\cal A}\cdot\varepsilon$, are listed in Table~\ref{tab:yields}. 
The number of visible single tops in $t\,W$ processes is smaller by a factor of $\sim$30 compared to those
from $\ttbar$ production, due to a combination of causes: lower cross sections, smaller reconstruction
efficiencies, and only one top-quark per event. At the LHC, we expect about 100 and 300 top-quarks measurable
in the fully leptonic decays from $\ttbar$-pairs and $t\,W$ processes in \PbPb\ and \pPb\ respectively. 
For comparison, the CDF and D0 experiments reconstructed less than 100 top-quarks (in all decays channels) during the full Run-1 operation at Tevatron.
At the end of the LHC heavy-ion programme, with $\LumiInt\approx$~10~nb$^{-1}$ (1~pb$^{-1}$) integrated in \PbPb\ (\pPb), about 
2.5 thousand (fully-leptonic) (anti)top-quarks should have been measured individually by the CMS and ATLAS
experiments. The corresponding visible yields at the FCC are about 300 times larger, reaching 5$\times10^4$ and 10$^5$ top-quarks per
year in \PbPb\ and \pPb\ collisions respectively. 

\begin{table}[htbp]
\caption{Expected number of top+antitop quarks per run, after typical acceptance and efficiency losses, 
for top-pair and $t\,W$ single-top 
measurements in fully-leptonic final-states in \pPb\ and \PbPb\ collisions at LHC and FCC energies for the
nominal per-year luminosities quoted.
\label{tab:yields}}
\vspace{-0.25cm}
\begin{center}
\begin{tabular}{lcc|c|c}\hline\hline
\hspace{0mm} System \hspace{0mm} & \hspace{0mm} $\sqrts$ \hspace{0mm} & \hspace{0mm} $\LumiInt$ \hspace{0mm} 
& \hspace{0mm} Number of top+antitop quarks \hspace{0mm} & \hspace{0mm} Number of top+antitop quarks  \hspace{0mm} \\
&  & & \hspace{0mm} $\ttbar\to \bbbar\,\ell\ell\,\nu\nu$ \hspace{0mm} & \hspace{0mm} $t\,W \to b\,\ell\ell\,\nu\nu$ \hspace{0mm} \\\hline
\hspace{0mm} \PbPb & 5.5~TeV \hspace{0mm} & \hspace{0mm} 1 nb$^{-1}$ \hspace{0mm} & \hspace{0mm} 90 \hspace{0mm} & \hspace{0mm} 3 \hspace{0mm} \\
\hspace{0mm} \pPb  & 8.8~TeV \hspace{0mm} & \hspace{0mm}0.2 pb$^{-1}$\hspace{0mm} & \hspace{0mm} 300 \hspace{0mm} & \hspace{0mm} 10 \hspace{0mm} \\\hline

\hspace{0mm} \PbPb & 39.~TeV \hspace{0mm} & \hspace{0mm} 5 nb$^{-1}$ \hspace{0mm} & \hspace{0mm} 47\,000\hspace{0mm} & \hspace{0mm}1\,300 \hspace{0mm} \\
\hspace{0mm} \pPb  & 63.~TeV \hspace{0mm} & \hspace{0mm} 1 pb$^{-1}$ \hspace{0mm} & \hspace{0mm} 100\,000\hspace{0mm} & \hspace{0mm}2\,600 \hspace{0mm} \\
\hline\hline

\end{tabular}
\end{center}
\end{table}

\begin{figure}[htpb]
\centering
\epsfig{figure=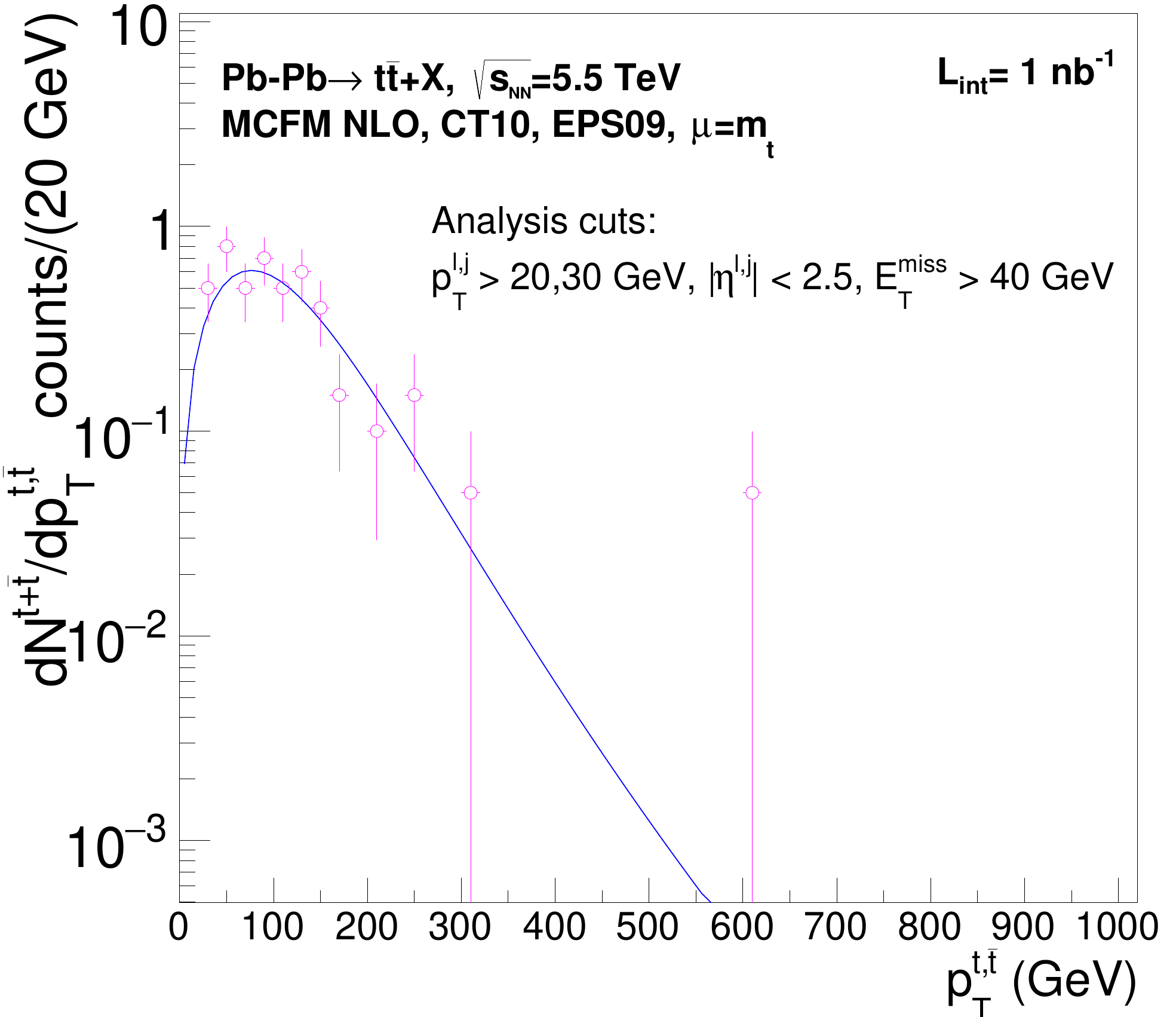,width=0.45\columnwidth}
\epsfig{figure=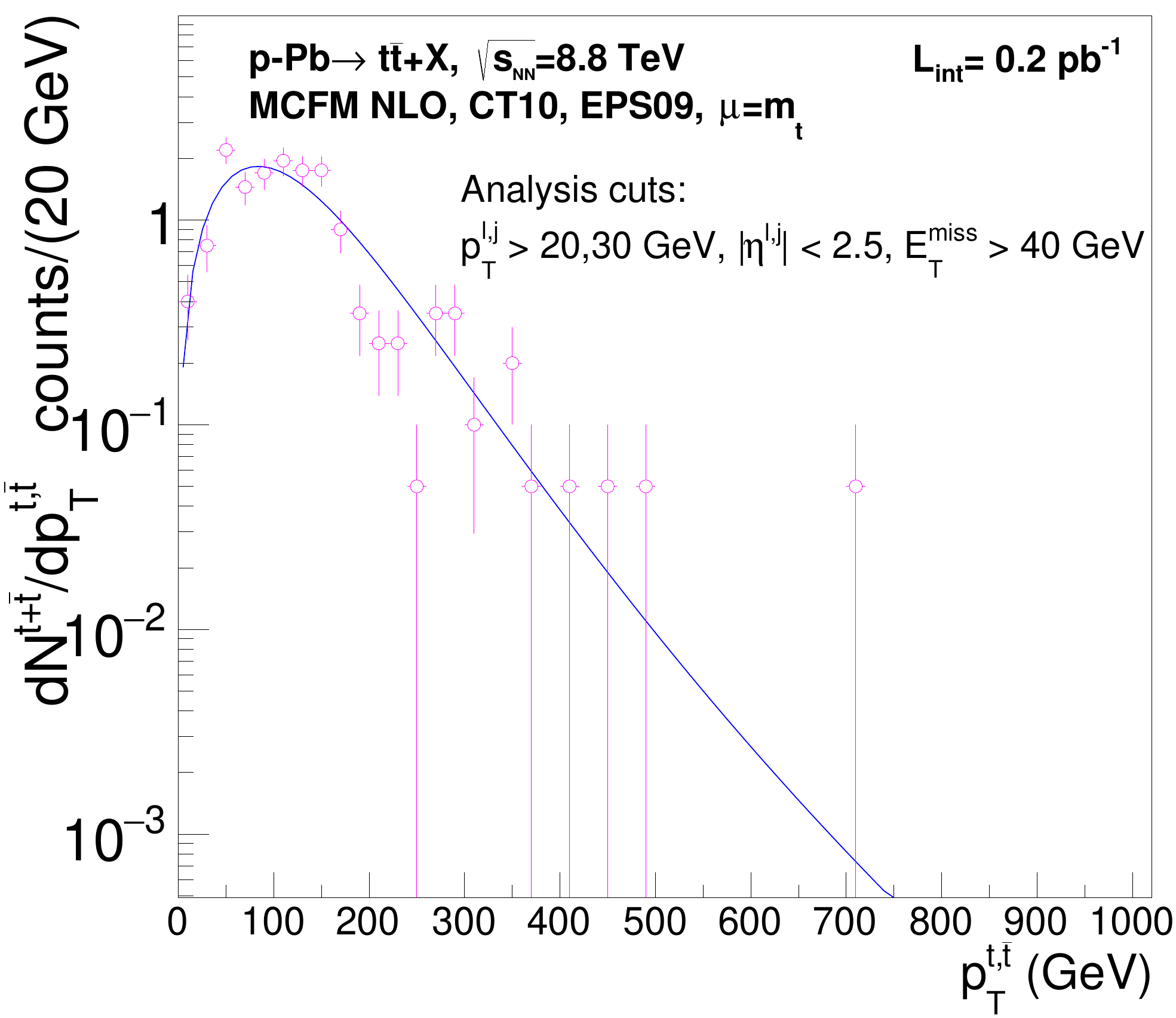,width=0.45\columnwidth}
\epsfig{figure=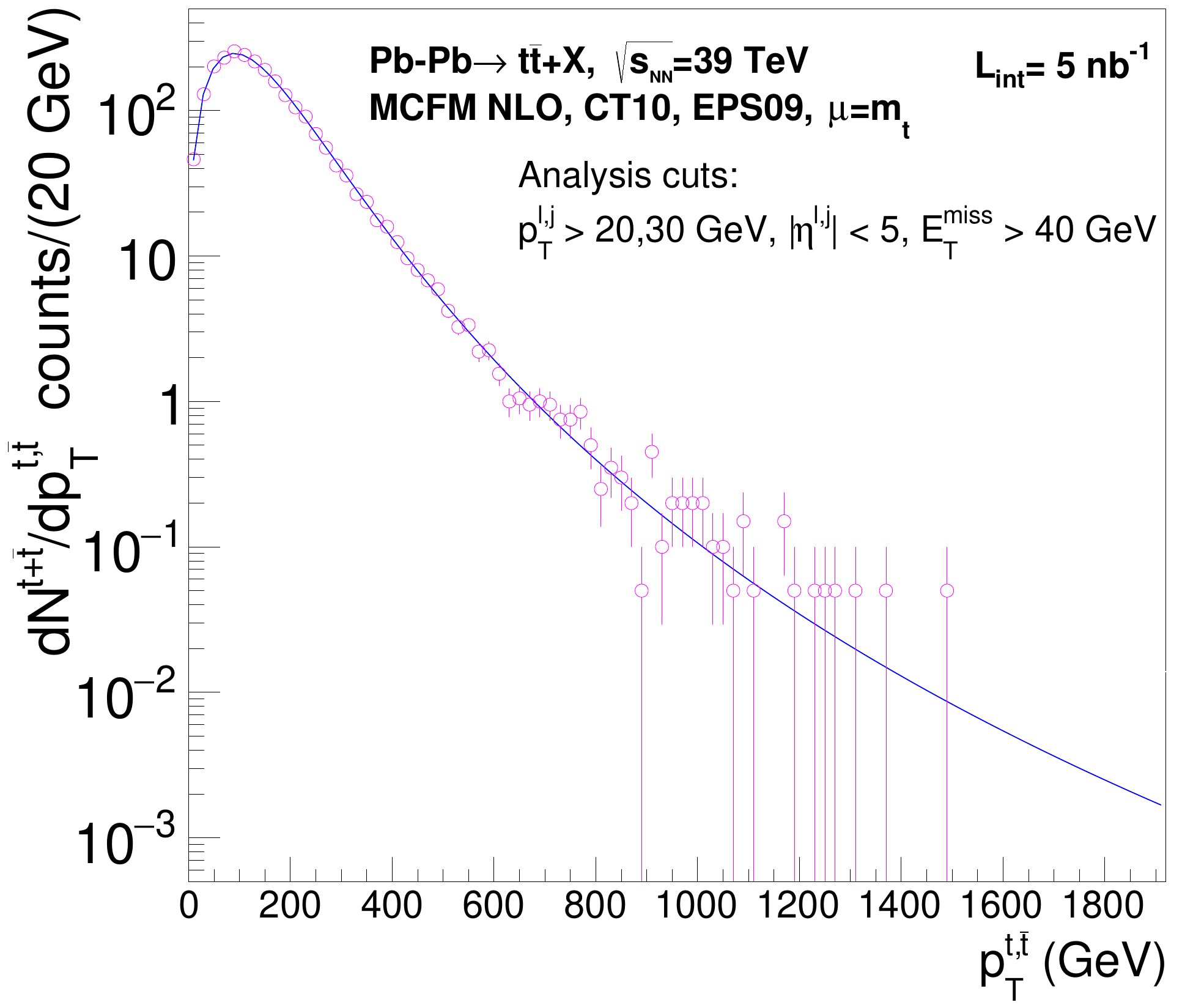,width=0.45\columnwidth}
\epsfig{figure=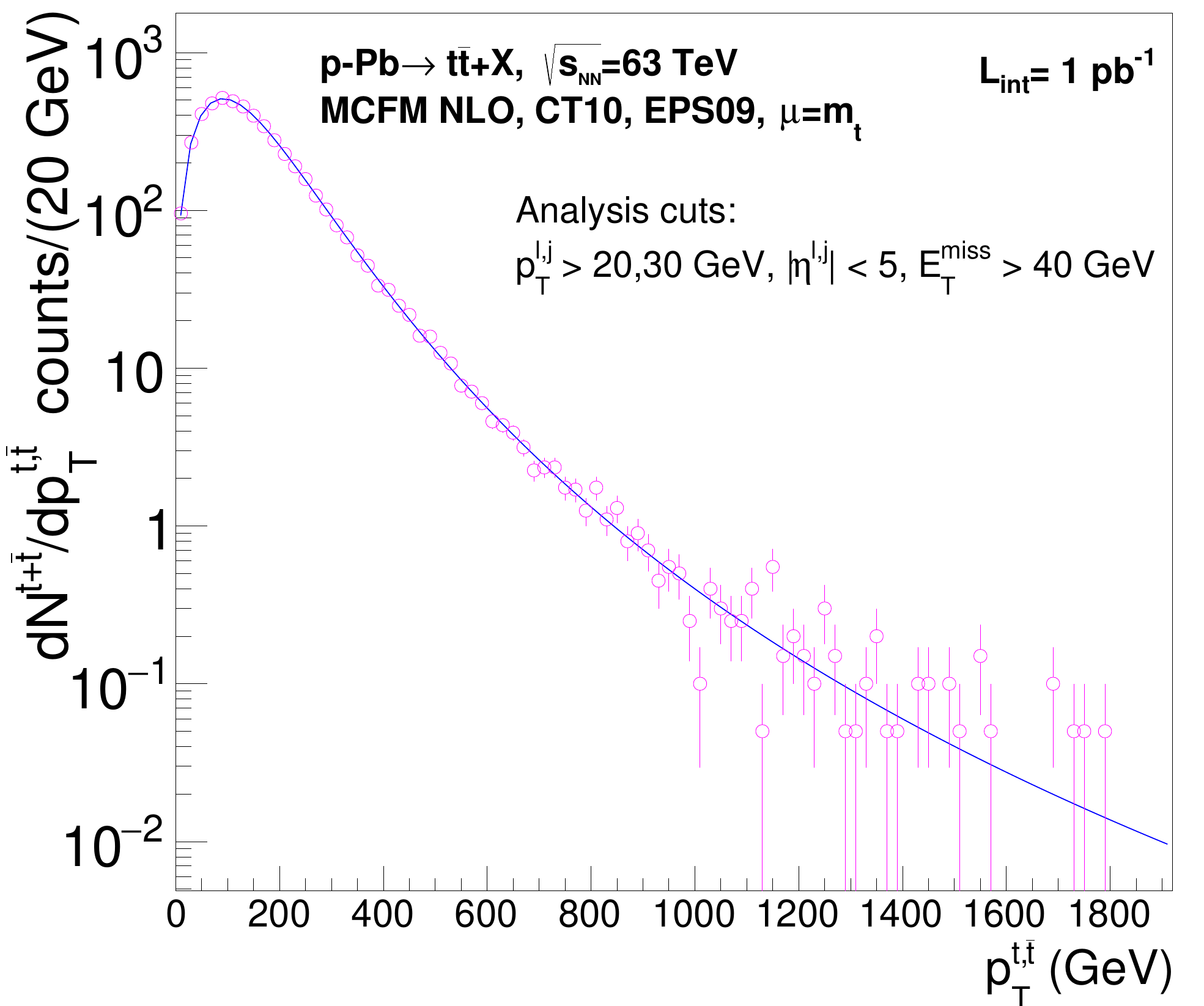,width=0.45\columnwidth}
\caption{Expected top-quark $\pt$ distributions, $dN^{t+\overline{t}}/dp_{\rm T}^{t,\overline{t}}$, in \PbPb\ (left panels) and \pPb\ (right panels) in the
  fully-leptonic decay modes at $\sqrtsnn$~=~5.5,\,8.8~TeV (LHC, top panels) and 39,\,63~TeV (FCC, bottom
  panels) after acceptance and efficiency cuts. 
  The curves are a fit to the underlying \mcfm\ distribution. The markers indicate pseudodata
  corresponding to the luminosities listed in Table~\ref{tab:yields}.
\label{fig:ptreach}}
\end{figure}

In order to provide an idea of the top-quark $\pt$ reach accessible in the different measurements listed in
Table~\ref{tab:yields}, we have also computed 
the expected $t,\bar{t}$ transverse-momentum
distributions in \PbPb\ and \pPb\ collisions, after all acceptance and efficiency criteria applied
(Fig.~\ref{fig:ptreach}).
The maximum top-quark $\pt$ experimentally measurable per LHC (FCC) year will be around $\pt\sim$~300
(1500)~GeV for \PbPb, and $\pt\sim$~500 (1800)~GeV for \pPb. Given the limited LHC top-quark statistics, the
study of boosted-tops will be thus only accessible at the future circular collider.

\section{Constraints on nuclear PDFs from $t\overline{t}$ production}
\label{sec:4}

As aforementioned, 80--95\% of the total pair production at LHC--FCC comes from gluon-gluon fusion
processes and, thus, $\ttbar$ cross sections can be used to constrain the relatively badly-known gluon
densities in the Pb nucleus. 
In this section we quantify the impact that top-quark measurements at the LHC and FCC would have
on better constraining the nuclear PDFs through the so-called Hessian PDF reweighting
technique~\cite{Paukkunen:2013grz,Paukkunen:2014zia}.
Such PDF reweighting procedure is based on the fact the error sets $\{f\}_k^\pm$ defined in Hessian PDF-fits 
correspond to a certain increment $\Delta \chi^2$ ( $\Delta \chi^2=50$ for EPS09) of the global ``goodness-of-fit''
$\chi^2$ function whose minimum $\chi^2_0$ is achieved with the central set $\{f\}_0$. The error sets thereby
constitute a parametrization of the original $\chi^2$ function which can be taken advantage of in order to determine
the associated PDF uncertainty after adding new experimental datasets. More precisely, the Hessian method
\cite{Pumplin:2001ct} for determining the PDF errors writes the response of the original $\chi^2$ function to
fit-parameter variations $\delta a_i$ as  
\begin{equation}
\chi^2\{a\} - \chi^2_0 \approx \sum_{i,j} \delta a_i H_{ij} \delta a_j = \sum_k  z_k^2,
\label{eq:chi2original}
\end{equation}
where the Hessian matrix $H_{ij}=(1/2) {\partial^2 \chi^2} / ({\partial{a_i}\partial{a_j}})$ is diagonalized
in the last step. The PDF error sets $\{f\}_k^\pm$ are then defined by $z_i({\{f\}_k^\pm})= \pm \sqrt{\Delta \chi^2} \delta_{ik}$.
The impact of including new experimental data can now be computed by considering a function
\begin{equation}
\chi^2_{\rm new} \equiv  \sum_k z_k^2  + \sum_{i,j} \left[T_i(z)-D_i\right] C_{ij}^{-1} \left[T_j(z)-D_j\right],
\label{eq:newchi2} 
\end{equation}
where $D_i$ denote the $i$th new data-point and $C$ is the covariance matrix that encodes the experimental
uncertainties. The corresponding theoretical values are denoted by $T_i$ and they depend on the PDFs. 
To first approximation, the $z$ dependence of each $T_i$ is given by
\begin{equation}
 T_i \left(z \right) \approx T_i \left(0 \right) + \sum_{k} \frac{T_i(z_k^+) - T_i(z_k^-)}{2} \, w_k,
\label{eq:T}
\end{equation}
where $w_k \equiv {z_k}/{\sqrt{\Delta \chi^2}}$, $T_i \left(0 \right)$ is the theory value computed 
with the central set $\{f\}_0$, and $T_i(z_k^\pm)$ is the theory value evaluated with the error set $\{f\}_k^\pm$.
Inserting Eq.~(\ref{eq:T}) into Eq.~(\ref{eq:newchi2}) and requiring $\partial \chi^2_{\rm new}/\partial w_k=0$
for all $k$, one finds the condition ${{w}^{\rm min}_k} = -\sum_i {B}^{-1}_{ki} {a_i}$ for the minimum of $\chi^2_{\rm new}$,
where 
$
 B_{kn} = \sum_{i,j} D_{ik} C_{ij}^{-1} D_{jn}  + \Delta \chi^2 \delta_{kn} \,$, 
$a_k   =  \sum_{i,j} {D_{ik} C_{ij}^{-1} \left[ T_j\left(0\right) - D_j \right]},$ and 
$D_{ik}  = \left(T_i(z_k^+ \right) - T_i\left(z_k^- )\right)/2.$ 
The new theory values $T_i^{\rm new}$ are obtained directly from Eq.~(\ref{eq:T}) with $w_k=w_k^{\rm min}$, and
the corresponding new central set of PDFs $\{f^{\rm new}\}_0$ by replacing $T_i$ with the PDFs,
$T_i \rightarrow f^{\rm new}(x,Q^2)$. To find the new PDF error sets $\{f^{\rm new}\}_k^\pm$, 
we rewrite Eq.~(\ref{eq:newchi2}) as
\begin{equation}
 \chi^2_{\rm new} - \chi^2_{\rm new}{}_{\big|{\bf w}={\bf w}^{\rm min}}  = 
 \sum_{i,j} \delta w_i B_{ij} \delta w_j  = \sum_k v_k^2,
\label{eq:newchi2other}
\end{equation}
where $\delta w_i = w_i-w_i^{\rm min}$, and in the last step the matrix $B$ is being diagonalized. The new PDF error sets
$\{f^{\rm new}\}_k^\pm$ can be then defined exactly as earlier by $v_i({\{f^{\rm new}\}_k^\pm}) = \pm \sqrt{\Delta \chi^2} \delta_{ik}$.
The procedure sketched above was proven in Ref.~\cite{Paukkunen:2014zia} to be consistent with the Bayesian reweighting
method introduced originally in Ref.~\cite{Giele:1998gw} and further confirmed more recently in
Ref.~\cite{Sato:2013ika}.


To mimic a realistic experimental situation, we generate sets of pseudodata for nuclear-modification 
factors $R_{_{\rm pPb}}$ and $R_{_{\rm PbPb}}$ --i.e. for the ratios of cross sections obtained
with EPS09 
nPDFs over those obtained with the CT10 proton PDFs-- 
corresponding to the LHC and FCC scenarios discussed in
the previous Section. As noted earlier, the {\it total} $t\overline{t}$ production cross section is expected to
undergo only a mild increase due to nuclear effects in the nPDFs, and it may be challenging to resolve
such an effect from overall normalization uncertainties. 
Thus, the total $t\overline{t}$ cross sections are not expected to have as large impact as they have on the
absolute free proton PDFs~\cite{Czakon:2013tha}. In order to get better constraints on the nPDFs 
differential cross sections are thus needed. For this purpose, we use the distributions of leptonic top-decay products,
which are unaffected by final-state interactions in the strongly-interacting matter produced in nuclear
collisions. Here, we concentrate on the measurement of $t\overline{t}$ pairs via the $t\overline{t}\rightarrow
b\overline{b} + \ell^+ \ell^- + \nu\overline{\nu}$ decay channel, binned in the charged lepton rapidity
$dN_{\ell}/dy_{\ell}$ with $\ell$ = e, $\mu$.
The baseline for the pseudodata is taken from the nuclear modification factors computed with the
central set of EPS09 in the previous Section. The expected number of events $N\left(\Delta y_i \right)$ in
each rapidity bin $\Delta y_i$ are computed by 
\begin{equation}
N\left( \Delta y_i \right) = N_{\rm total} \times \frac{\sigma(y_i \in \Delta y)}{\sigma_{\rm total}},
\end{equation}
where $N_{\rm total}$ is the total number of events expected after acceptance and efficiency losses for each
system\footnote{For the LHC, we consider the total luminosity to be accumulated during the full heavy-ion
programme, which is a factor of 10 (5) higher than the nominal per-year luminosities quoted for \PbPb\ (\pPb).} 
listed in Table~\ref{tab:yields}, $\sigma(y_i \in \Delta y)$ is the cross section within rapidity
bin $\Delta y_i$, and $\sigma_{\rm total}$ is the total cross section within the acceptance. 
The statistical uncertainty is then taken to be $\delta_i^{\rm stat} = T_i^{_{\rm EPS09}} \sqrt{1/N\left( \Delta y_i \right)}$
to which we add in quadrature a constant $\pm$5\% systematic error  ($\delta_i^{\rm syst}=0.05 \times T_i^{_{\rm EPS09}}$), 
such that the total uncorrelated error is $\delta_i^{\rm uncorr} = \sqrt{(\delta_i^{\rm stat})^2 + (\delta_i^{\rm syst})^2}$.
The overall normalization error is taken to be 5\% ($\delta_i^{\rm norm} = 0.05 \times T_i^{_{\rm EPS09}}$).
Systematic uncertainties of this order are realistic as the corresponding \pp\
measurements at the LHC~\cite{Chatrchyan:2012bra,Chatrchyan:2012saa} have already reached
a better precision. In addition, partial cancellations of systematic uncertainties common to \pp, \pPb,
and \PbPb\ measurements are expected in a careful experimental determination of $R_{_{\rm pPb}}$ and $R_{_{\rm PbPb}}$.
The statistical precision of the \pp\ baseline data, taken at slightly different center-of-mass energies,
is also much better than that of the \pPb\ and \PbPb\ measurements, and the additional theory uncertainty for energy-dependent
corrections of the reference cross sections (e.g. to go from $\sqrts$~=~8~TeV in \pp\ to $\sqrtsnn$~=~8.8~TeV in
\pPb\ at the LHC) are small~\cite{Mangano:2012mh}.
Each pseudodata point $D_i$ is then computed from the baseline values $T_i^{_{\rm EPS09}}$ and from the
estimated uncorrelated and normalization errors by 
$
D_i = \left( T_i^{_{\rm EPS09}} + \delta_i^{\rm uncorr} r_i + \delta_i^{\rm norm} r^{\rm norm} \right),$ 
where $r_i$ and $r^{\rm norm}$ are random numbers from a Gaussian distribution of variance one centered around zero.
The elements of the covariance matrix $C$ are computed as 
$
C_{ij} = \left[ \delta_i^{\rm uncorr}\delta_j^{\rm uncorr}  \delta_{ij} + \delta_i^{\rm norm}\delta_j^{\rm norm}  \right] 
$~\cite{Albino:2008fy}.

\begin{figure}[htb!]
\begin{minipage}{\textwidth}
\includegraphics[width=0.49\textwidth]{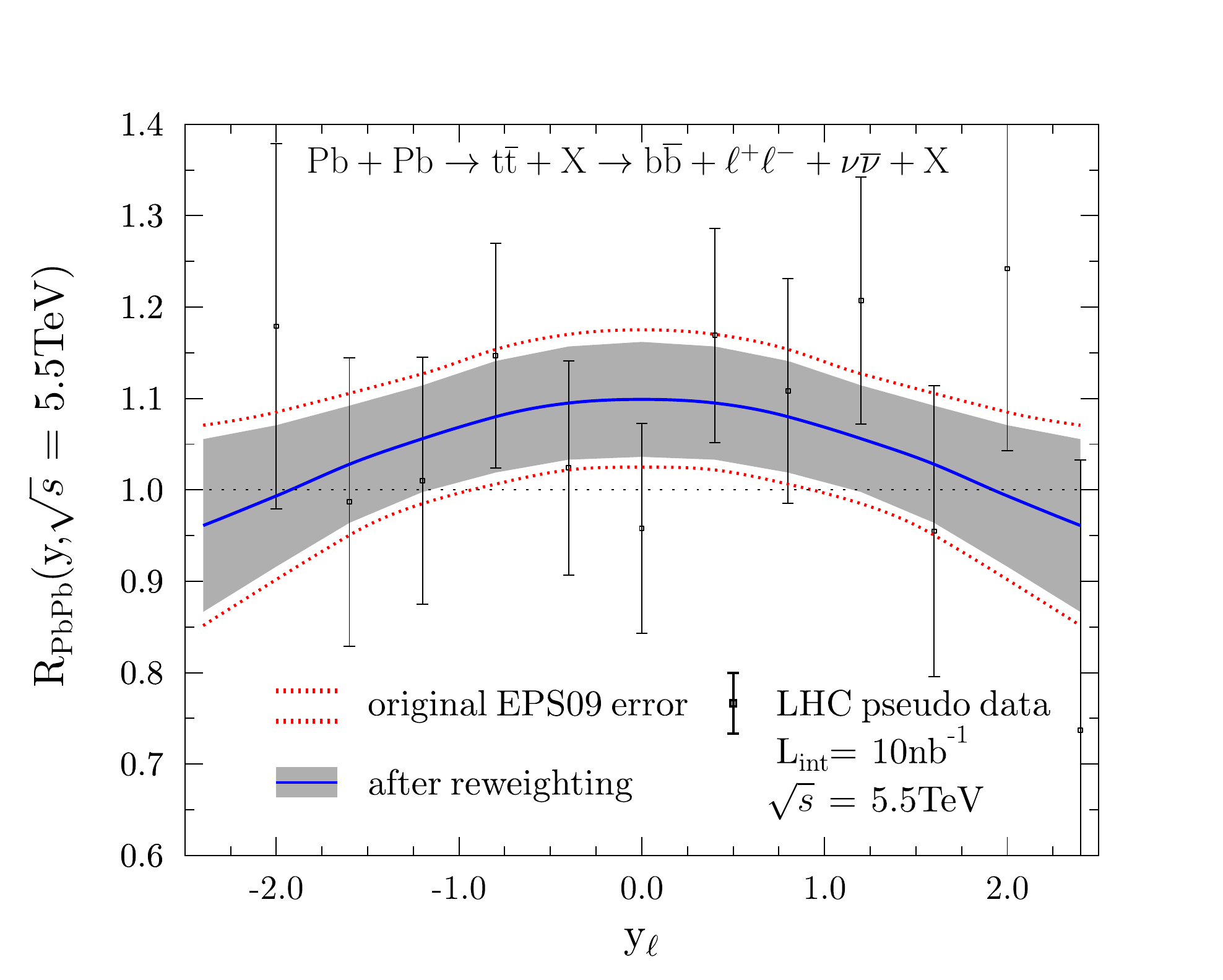}
\includegraphics[width=0.49\textwidth]{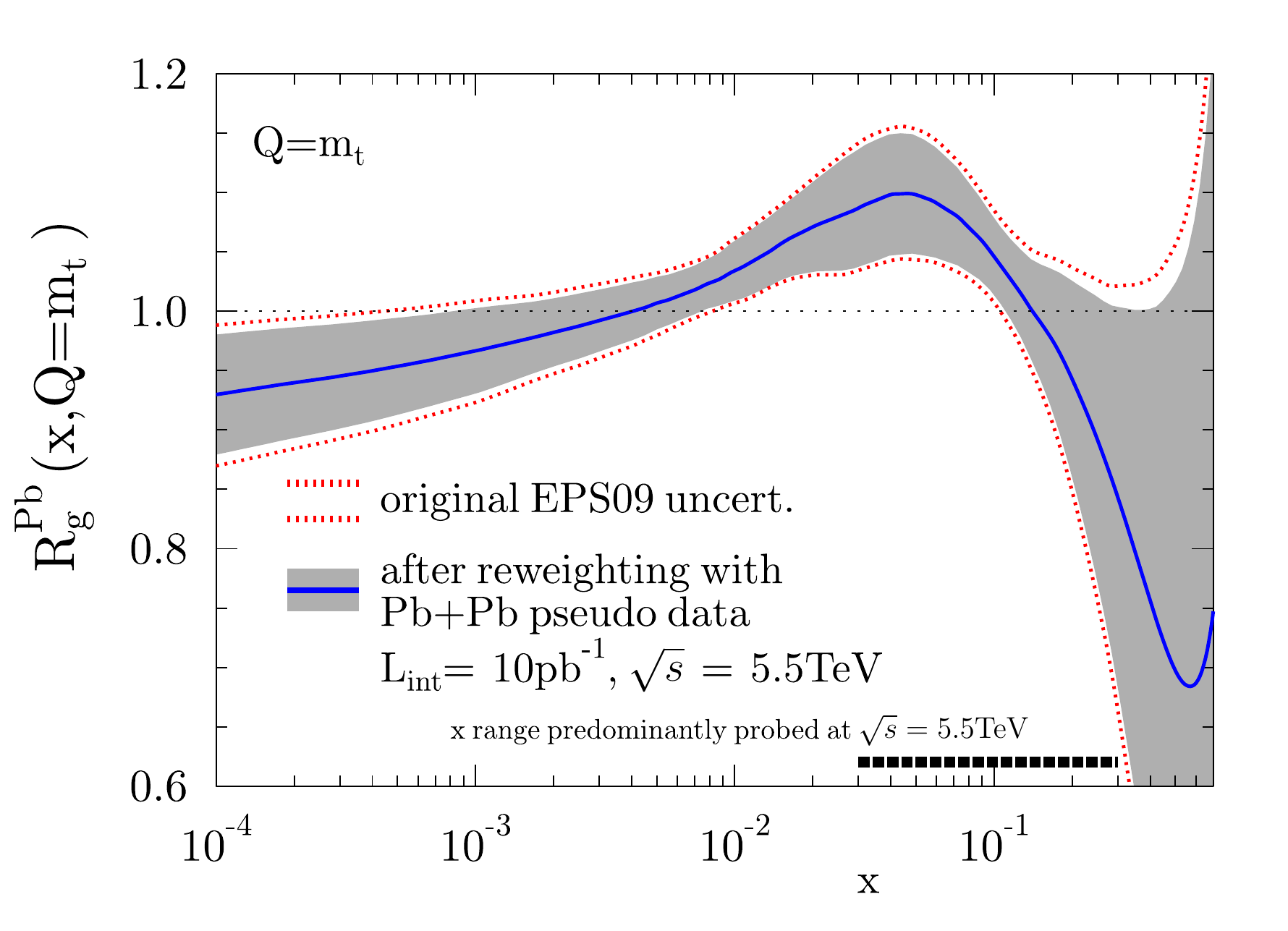}
\includegraphics[width=0.49\textwidth]{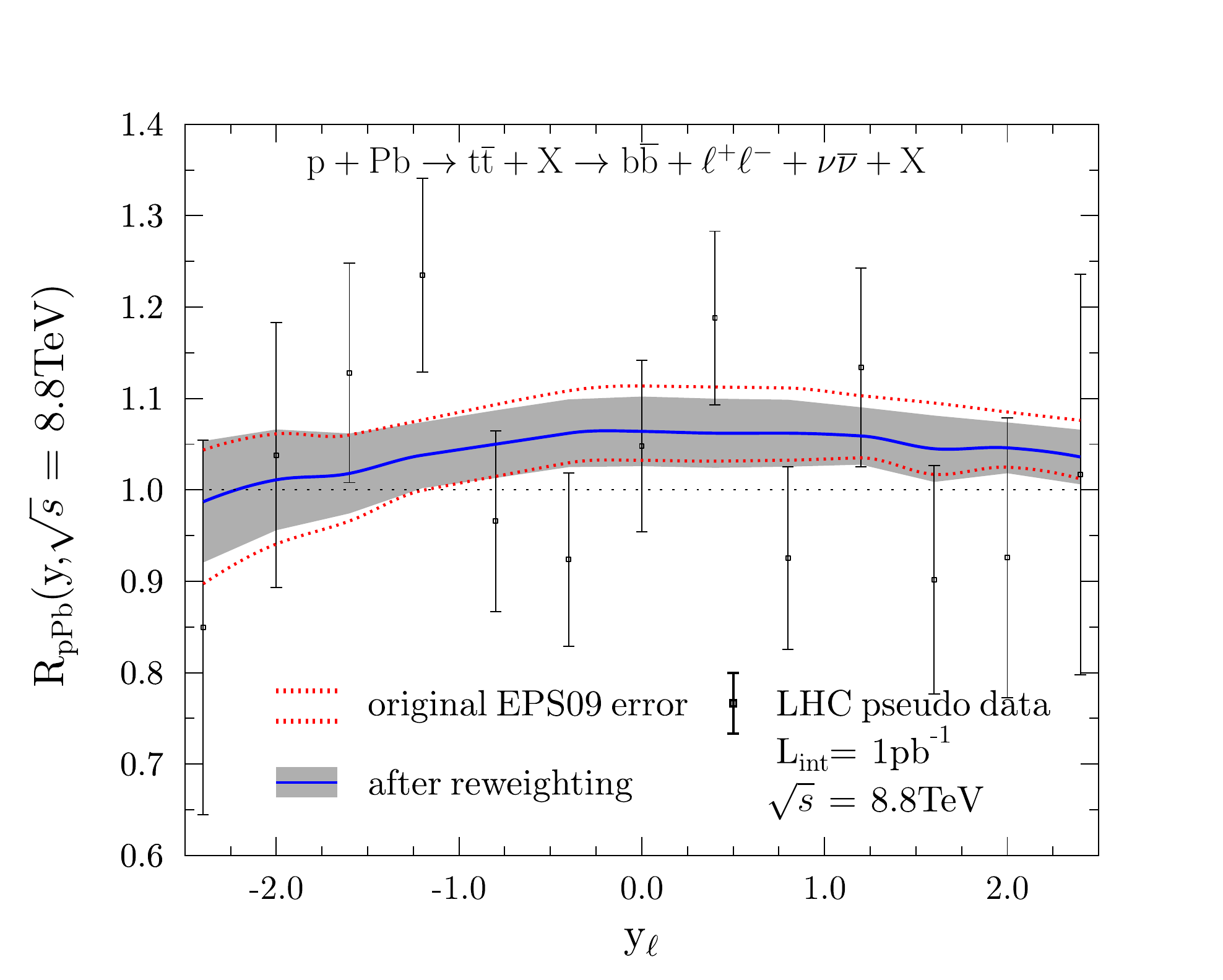}
\includegraphics[width=0.49\textwidth]{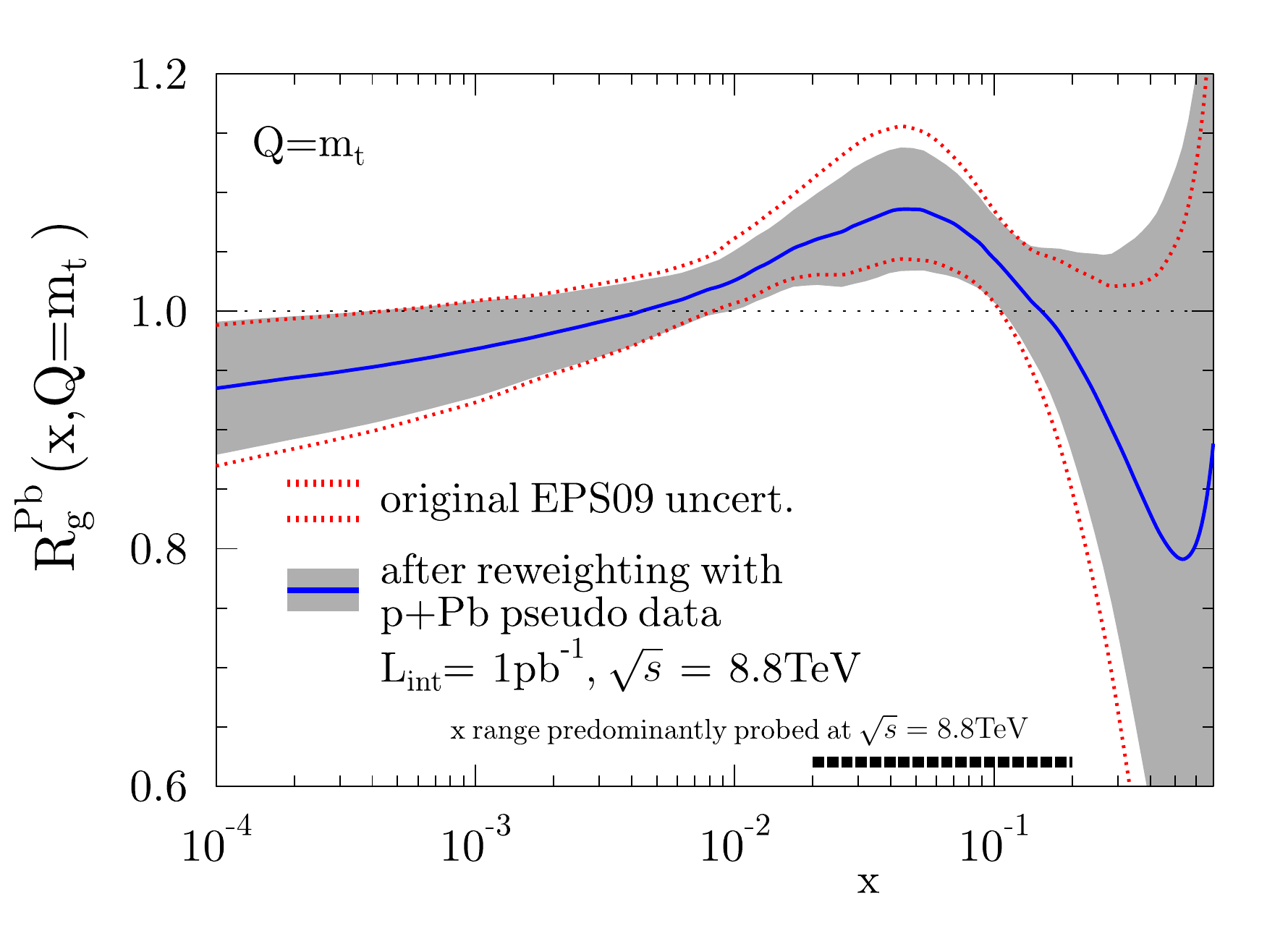}
\caption{Impact on the nuclear glue of $\ttbar$ (pseudo)data for the full LHC heavy-ion programme.
  Left: Nuclear modification factor for $\ttbar$ decay-leptons as a function of rapidity, R$_{_{\rm PbPb}}(y_{\ell})$,
  obtained for \PbPb\ at $\sqrtsnn$~=~5.5~TeV (top) and \pPb\ at $\sqrtsnn$~=~8.8~TeV (bottom)
  compared to predictions computed with  
  EPS09: current nPDF set (region enclosed by the red dotted lines) and after
  pseudodata-reweighting 
  (blue curve plus grey band).
  Right: Ratio of nuclear-over-proton gluon densities, R$_{_{\rm g}}^{^{\rm Pb}}$ evaluated at 
  $Q=m_{\rm t}$, for the original EPS09 
  (band enclosed by red dotted lines) and for the reweighted EPS09 (blue curve with grey band) 
  for \PbPb\ (top) and \pPb\ (bottom).}
\label{fig:data1}
\end{minipage}
\end{figure}

In the original EPS09 analysis~\cite{Eskola:2009uj}, the inclusive pion data measured at RHIC~\cite{Adler:2006wg}
was given an additional weight factor of 20 in the $\chi^2$-function in order to enhance the constraints on
the badly known gluon densities from nuclear deep-inelastic and fixed-target Drell-Yan data alone. As both the pion data and the top
quark production considered here, are mostly sensitive to nuclear gluon PDFs, we rescale the covariance matrix
equally by $C \rightarrow C/20$ when performing the reweighting. This compensates for
the large weight given for the RHIC pion data and should lead to a more realistic estimate of the impact that the
top-quark measurements would have if directly included into the EPS09 fit. After finding the new theory values
$T_i^{\rm new}$ through the reweighting 
procedure, the optimum overall shift (originating from the allowed uncertainty in normalization) is found by
solving the multiplicative factor $f$ from the $\chi^2$ contribution of the new data (see e.g. Ref.~\cite{Albino:2008fy}):
\begin{equation}
\sum_{i,j} \left[T_i^{\rm new}-D_i\right] C_{ij}^{-1} \left[T_j^{\rm new}-D_j\right] = \min \left\{
\sum_i \left[ \frac{T_i^{\rm new} - D_i - f \delta_i^{\rm norm}}{\delta_i^{\rm uncorr}} \right]^2 + f^2 \right\}.
\label{eq:chi2syst}
\end{equation}
In the results presented below, the resulting shift $f \delta_i^{\rm norm}$ has been applied
on the data points.


The results of the nPDF reweighting procedure are presented first for LHC energies
in Figure~\ref{fig:data1} for \PbPb\ (top) and \pPb\ (bottom) collisions. 
The nuclear modifications for the decay leptons are somewhat less pronounced in comparison to the 
top distributions themselves (Fig.~\ref{fig:dsigmady_ttbar}) due to the smoothing brought about
by the additional phase-space integrations related to the top-quark decays.
The estimated statistical errors are generally
of the order of 10\% and are foreseen to dominate over the predicted systematic uncertainties. 
These pseudodata probe the nuclear PDFs predominantly in the range  
$0.03 \lesssim x \lesssim 0.3$ at $\sqrt{s}=5.5 \, {\rm TeV}$, and $0.02 \lesssim x \lesssim 0.2$ at
$\sqrt{s}=8.8 \, {\rm TeV}$, as inferred from our \mcfm\ calculations.
In both cases the pseudodata are found to have only a moderate impact on the EPS09 gluon 
density\footnote{As the $Q^2$ dependence of $R_{\rm g}^{\rm Pb}(x,Q^2)$ is rather mild for $Q^2 \gtrsim 10 \,
  {\rm GeV}^2$ \cite{Eskola:2012rg}, the plots at $Q^2=m_{\rm t}^2$ are representative for most practical applications.}.
This is predominantly due to the rather low foreseen statistics and 
the rapidity interval covered, which makes especially $R_{_{\rm pPb}}(y_{\ell})$ somewhat flat within the acceptance. Consequently,
even the overall normalization alone can mimic the effects of nuclear PDFs thereby reducing the obtainable
constraints. The new error bands in Figs.~\ref{fig:data1} are both around 10\% narrower
than the original EPS09 ones. Combining the \pPb\ and \PbPb\ measurements and assuming independent data
samples available from both the CMS and ATLAS collaborations, the total impact of $\ttbar$
production on the large- and mid-$x$ nuclear gluons could reach 30\% (with the full LHC luminosity). Such a modest
improvement will most likely be overpowered by the constraints offered by the inclusive jet and
dijet data from the LHC \pPb\ run(s)~\cite{Paukkunen:2014pha}.



\begin{figure}[ht!]
\begin{minipage}{\textwidth}
\includegraphics[width=0.49\textwidth]{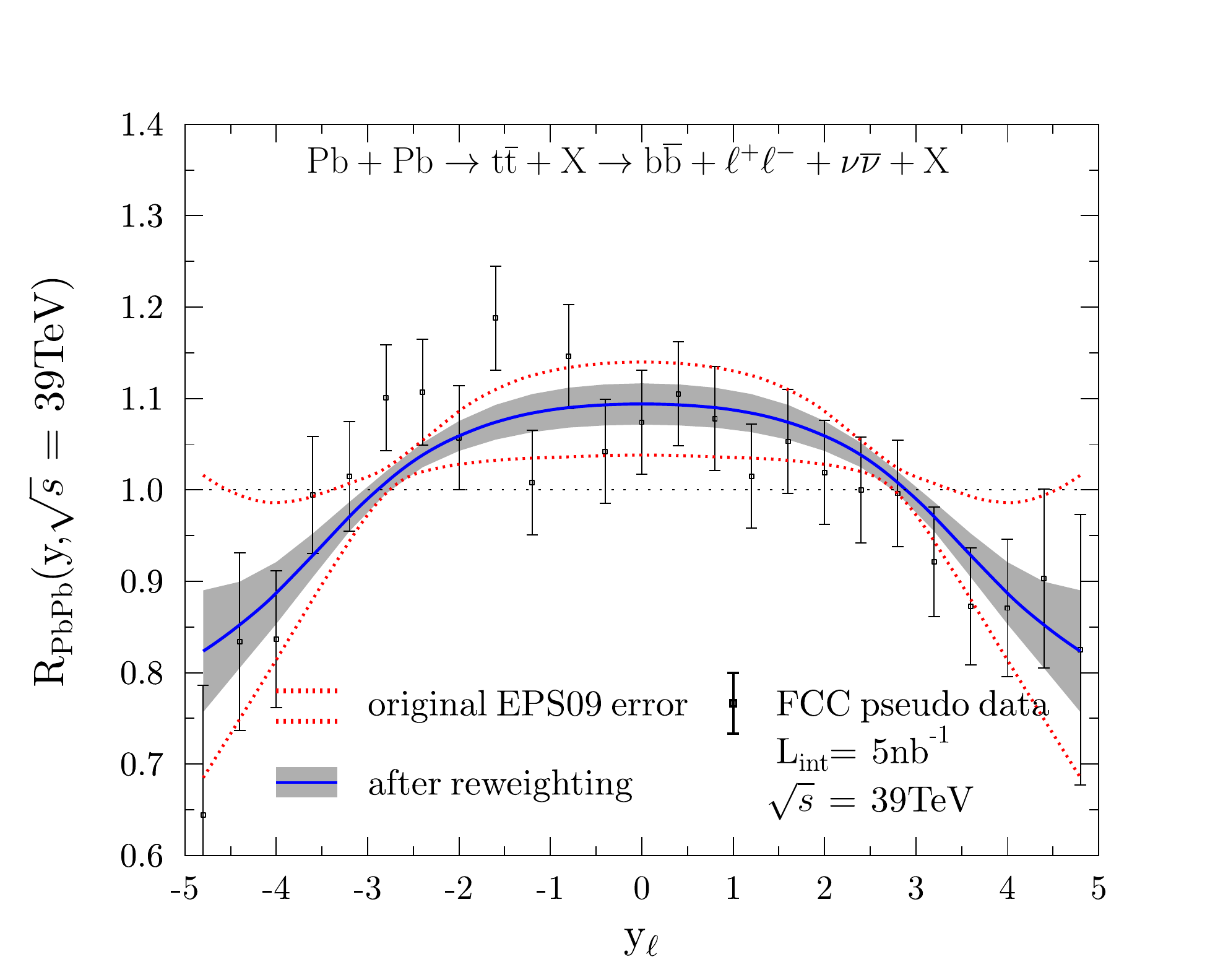}
\includegraphics[width=0.49\textwidth]{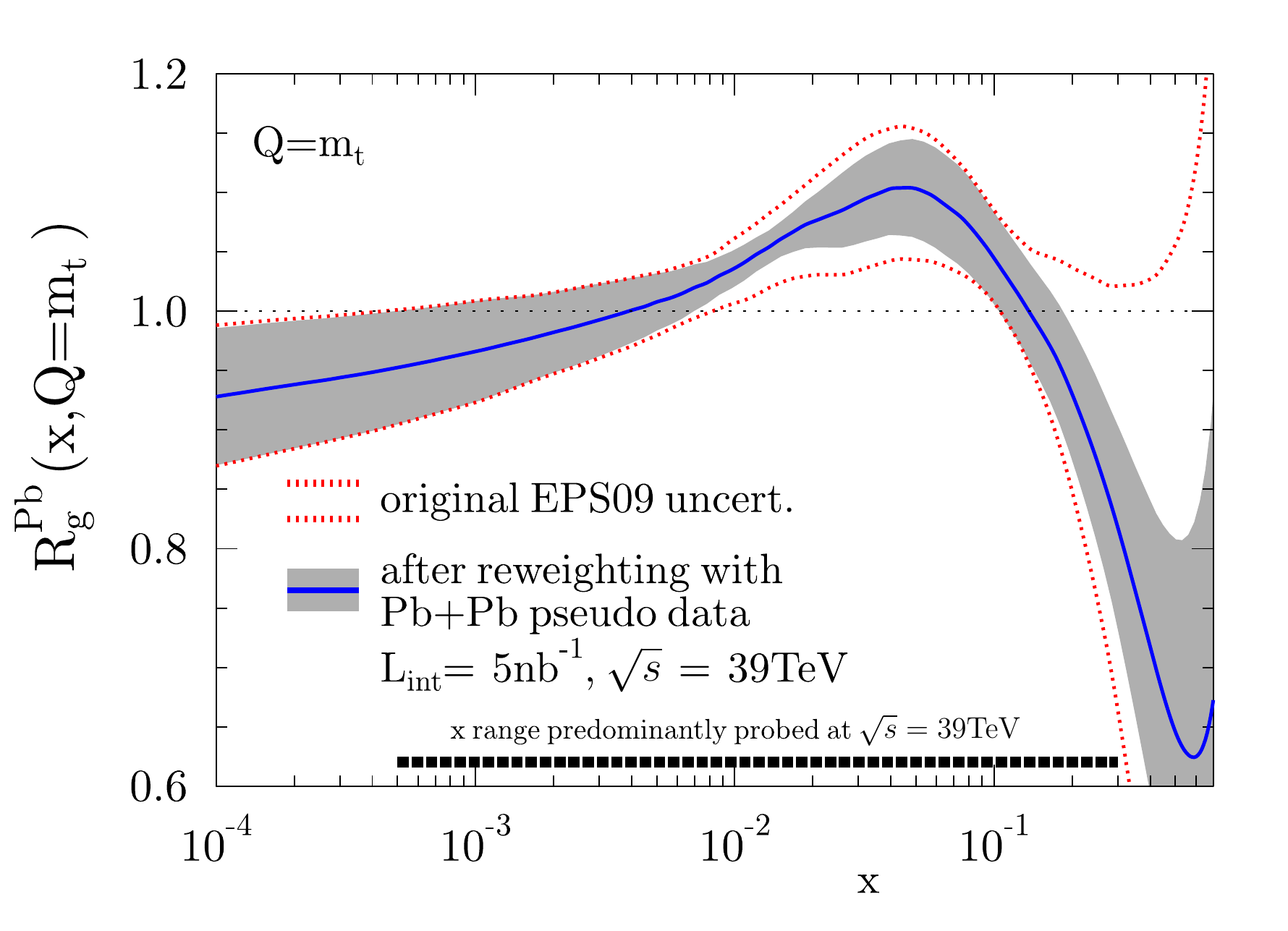}
\includegraphics[width=0.49\textwidth]{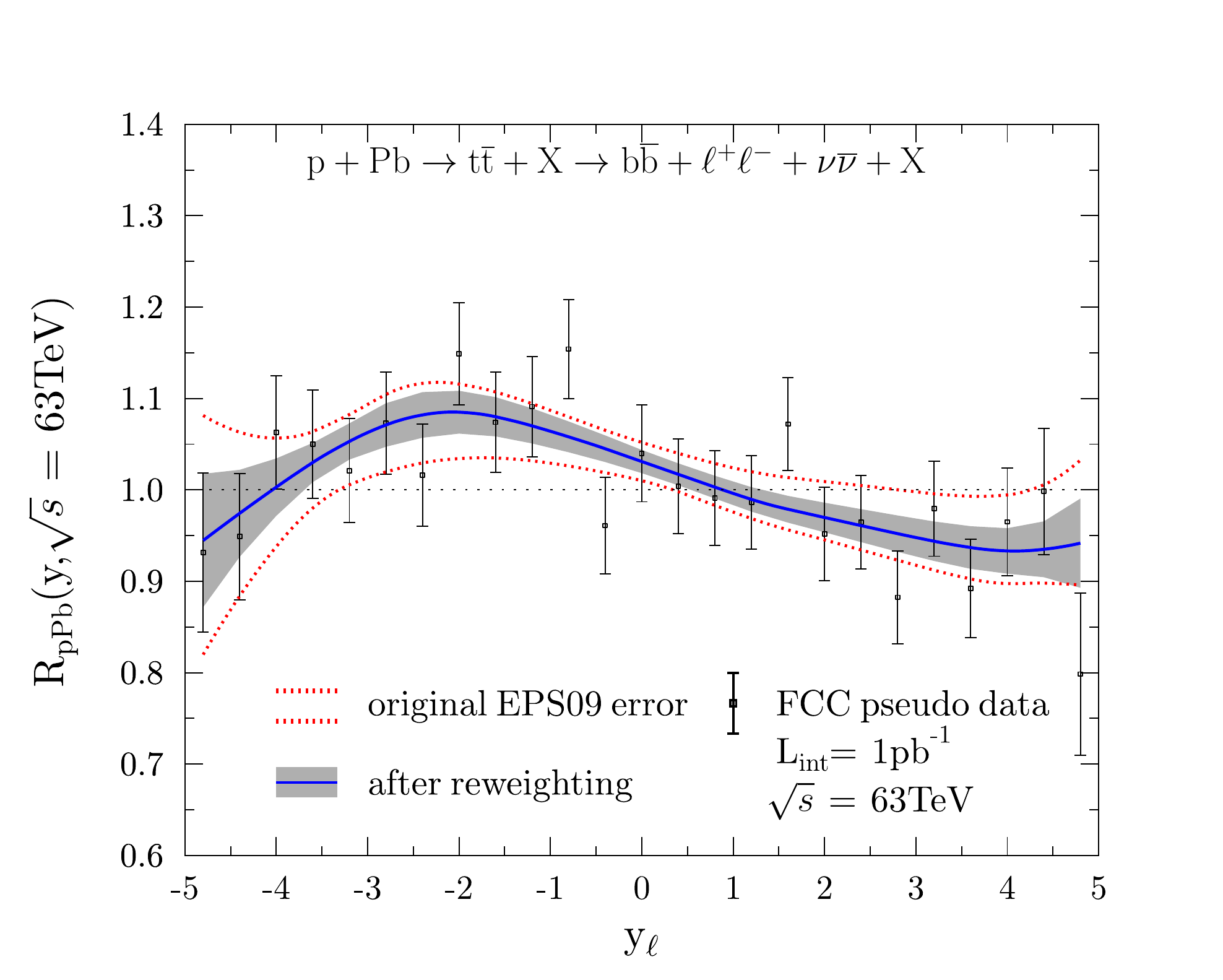}
\includegraphics[width=0.49\textwidth]{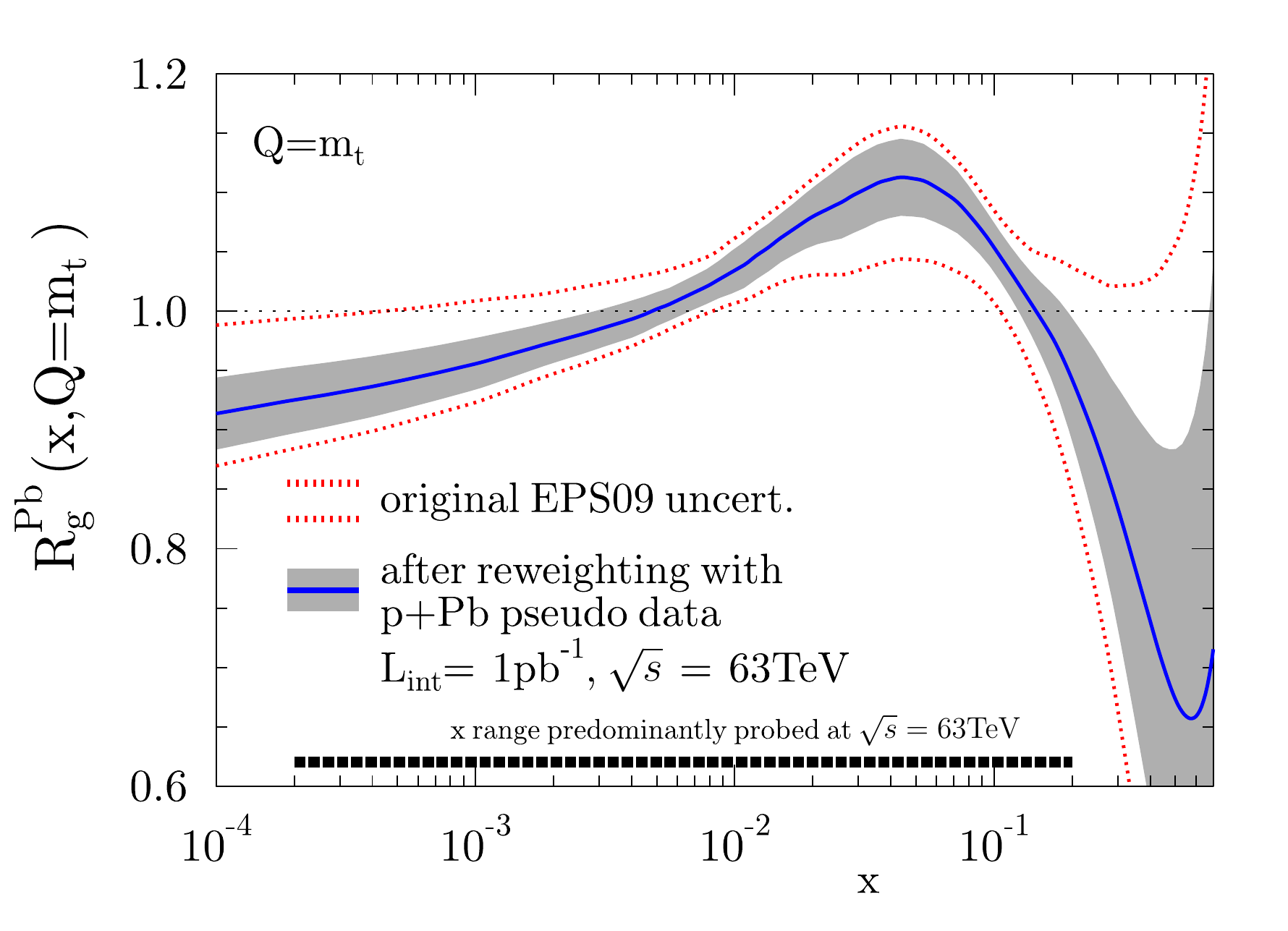}
\caption{As Figure~\ref{fig:data1} but for the case of \PbPb\ (top) and \pPb\ (bottom) per FCC-year pseudodata.} 
\label{fig:data3}
\end{minipage}
\end{figure}

The results of repeating the reweighting procedure with the FCC pseudodata are shown in
Fig.~\ref{fig:data3} for \PbPb\ (top) and \pPb\ (bottom) collisions respectively. 
Although the foreseen FCC per-year integrated luminosities are similar to those expected for the 
{\it full} LHC heavy-ion programme, the production cross sections being a factor $\times$(55--90) above the
LHC expectations (Table~\ref{tab:yields}) significantly increase the expected top-quark yields thereby
reducing the statistical uncertainties. Indeed, in our FCC scenario the systematic uncertainties dominate. 
The reduced uncertainties as well as the wider kinematic reach at the FCC make the impact of these pseudodata
on EPS09 clearly larger than that expected at the LHC. The constraints also reach lower values of $x$,
the dominant $x$ region being $5\times10^{-4} \lesssim x \lesssim 3\times10^{-1}$ at $\sqrt{s}=39 \, {\rm TeV}$, and
$2\times10^{-4} \lesssim x \lesssim 2\times10^{-1}$ at $\sqrt{s}=63 \, {\rm TeV}$.
The addition of the nuclear $\ttbar$ results shown in Fig.~\ref{fig:data3} would allow one to
decrease the gluon density uncertainty by up to 50\% in some regions. 
Unlike in \pPb\ collisions, in the \PbPb\ case the nuclear effects coming from
small-$x$ (shadowing) and large-$x$ (EMC effect) cannot be distinguished since
they are essentially multiplied at large $|y_\ell|$. In this case, the
new constraints tend to affect more the part that is originally less constrained,
the high-$x$ gluons. That is, while the probed region in $x$ is very similar
in our \pPb\ and \PbPb\ scenarios, the \pPb\ collisions are foreseen to provide more varied nPDF constraints.
Combination of \pPb\ and \PbPb\ data,
plus assumption of two independent experiments measuring the spectra, would result in an overall reduction of
up to 70\% with just one year of integrated luminosity.




\section{Conclusions}
\label{sec:5}

The study presented here has shown, for the first time, that top quarks produced in pairs via (mostly)
gluon-gluon fusion, or singly in electroweak processes, are clearly observable in \pPb\ and \PbPb\ collisions
at the energies of the CERN LHC and future circular collider (FCC). 
The corresponding cross sections have been computed at NLO accuracy with the \mcfm\ code including
the CT10 free proton PDFs and nuclear modifications parametrized with the EPS09 nPDF. 
At the LHC, the pair-production cross sections are $\sigma_{\ttbar}$~=~3.4~$\mu$b for \PbPb\ at
$\sqrtsnn$~=~5.5, and $\sigma_{\ttbar}$~=~50~nb for \pPb\ at $\sqrtsnn$~=~8.8~TeV.
At the FCC energies of $\sqrtsnn$~=~39,\;63~TeV, the same cross sections are factors of 90 and 55 times larger
respectively. The total $\ttbar$ cross sections are enhanced by 3--8\% in nuclear compared to \pp\ collisions
at the same c.m. energies, due to an overall net gluon antishadowing, although different regions of the 
top-quark differential distributions are depleted due to shadowing and EMC-effect corrections. 
The total cross sections for single-top, including the sum of $t$- and $s$-channels plus associated $t\,W$
processes, 
are a factor of two (four) smaller than that for top-pair production at the LHC (FCC)
and feature minimal nuclear modifications ($\pm$2\% depending on the energy).

After applying typical acceptance and efficiency cuts in the leptonic final-state, $\ttbar\to
W^+b\,W^-\bar{b}\to \bbbar\,\ell\ell\,\nu\nu$, one expects about 100 and 300 (anti)top-quarks per LHC-year
and 5$\times10^4$ and 10$^5$ per FCC-year at the nominal luminosities in \PbPb\ and \pPb\
collisions respectively.
At the end of the LHC heavy-ion programme, with $\LumiInt\approx$~10~nb$^{-1}$,\,1~pb$^{-1}$ integrated
\PbPb\ and \pPb\ luminosities, about 2.5 thousand (fully-leptonic) $t,\bar{t}$-quarks should have
been measured individually by the CMS and ATLAS experiments. 
The number of visible single-top quarks produced in association with a W boson, in the similar $t\,W \to
W\,b\,W \to b\,\ell\ell\,\nu\nu$ final state, is lower by a factor of about 30 compared to $\ttbar$ production,
due to the combination of lower cross sections, smaller reconstruction efficiencies, and only one
top-quark per event.


The proposed top-quark measurements at the LHC and FCC would not only constitute the first observation in nuclear collisions of the
heaviest-known elementary particle, but would open up interesting novel physics opportunities such as
constraints of nuclear parton densities in an unexplored kinematic range, 
studies of the dynamics of heavy-quark energy loss in the QGP, and colour-reconnection effects on the
top-quark mass. 
We have, in particular, quantified the impact on the nuclear PDFs of the rapidity-differential distributions
of the decay leptons from top-quark pairs, through the Hessian reweighting technique, finding that the data can be
used to reduce the uncertainty on the Pb gluon density at high virtualities by up to 30\% using the full LHC
heavy-ion programme, and by about 70\% with just one FCC-year. 


\section*{Acknowledgments}
We are grateful to Carlos Salgado for early discussions on this work, and to Andrea Dainese and Martijn Mulders for feedback on the
manuscript. K.K. acknowledges partial financial support from the Hungarian Scientific Research Fund (K 109703).



\end{document}